\documentclass[lettersize,journal]{IEEEtran}
\usepackage[utf8]{inputenc}
\usepackage[T1]{fontenc}
\usepackage[english]{babel}
\usepackage{array}
\usepackage{shortvrb}
\usepackage{listings}
\usepackage{amsmath}
\usepackage{amsfonts}
\usepackage{balance}
\usepackage{enumerate}
\usepackage{graphicx}             % import, scale, and rotate graphics
\usepackage[caption=false]{subfig}% group figures
\usepackage{subfloat}
\usepackage{alltt}
\usepackage{url}
\usepackage{xspace}
\usepackage{indentfirst}
\usepackage{eurosym}
\usepackage{wasysym}
\usepackage{color}
\usepackage{varwidth}
\usepackage{framed}
\usepackage{tcolorbox}
\tcbuselibrary{theorems}
\usepackage{xifthen}
\usepackage{multirow}
\usepackage{tikz}
\usetikzlibrary{positioning,decorations.pathreplacing,calc,fit,shapes.geometric, arrows,shapes.symbols,arrows.meta}
\usepackage{mathtools}
\usepackage{enumitem}
\usepackage{stackrel}
\usepackage{colortbl}
\usepackage{nth}
\usepackage{booktabs}
\usepackage{orcidlink}

\graphicspath{{.}{./Pictures/}}

\definecolor{mygreen}{rgb}{0,0.6,0}
\definecolor{mymauve}{rgb}{0.58,0,0.82}
\definecolor{mygray}{rgb}{0.5,0.5,0.5}
\definecolor{RougeLosange}{RGB}{255,42,6}
\definecolor{BleuBloc}{RGB}{0,156,253}
\definecolor{MauveBloc}{RGB}{131,54,188}
\definecolor{OrangeBloc}{RGB}{222,106,16}
\definecolor{MACRouge}{RGB}{255,38,0}
\definecolor{MACBleu}{RGB}{4,51,255}
\definecolor{MACVert}{RGB}{0,143,0}
\definecolor{MACOrange}{RGB}{255,147,0}
\definecolor{dline}{RGB}{255,38,0}
\definecolor{done}{RGB}{4,51,255}
\definecolor{todo}{RGB}{0,143,0}
\definecolor{lowerbound}{RGB}{255,147,0}
\definecolor{upperbound}{RGB}{255,64,255}
\definecolor{variter}{RGB}{146,144,0}
\definecolor{vardata}{RGB}{169,169,169}
\definecolor{varsize}{RGB}{148,23,81}
\definecolor{unmodified}{RGB}{0,150,255}
\definecolor{green-yellow}{rgb}{0.68, 1.0, 0.18}
\definecolor{red0946}{RGB}{255,38,0}
\definecolor{green0946}{RGB}{43,147,43}
\definecolor{Sepia}{HTML}{671800}

\newcommand{\dfn}[1]{\textit{#1}}            %introducing new terms

\newcommand{\fb}{feedback\xspace}
\newcommand{\ffw}{feedforward\xspace}
\newcommand{\Fb}{Feedback\xspace}

\newcommand{\gli}{Graphical Loop Invariant\xspace}
\newcommand{\glibpFull}{Graphical Loop Invariant Based Programming\xspace}

\newcommand{\bgli}{Blank \gli}

\newcommand{\cafe}{\textsc{Caf\'e} 2.0\xspace}

\newcommand{\cafeun}{\textsc{Caf\'e} 1.0\xspace}

\newcommand{\gamecode}{\textsc{GameCode}\xspace}
\newcommand{\gamecodes}{\textsc{GameCode}s\xspace}
\newcommand{\cdb}{\textsc{Cdb}\xspace}
\newcommand{\glibp}{\textsc{Glibp}\xspace}
\newcommand{\pca}{\textsc{Pca}\xspace}

\newcommand{\lv}{Loop Variant Function\xspace}%{Loop Variant\xspace}

\newcommand{\zone}[1]{\textsc{Initial Representation}\xspace} %{\textsc{Zone} #1\xspace}

\newcommand{\glide}{\textsc{Glide}\xspace}
\newcommand{\devphase}{\textit{concrete phase}\xspace}
\newcommand{\absphase}{\textit{abstraction phase}\xspace}

\newcommand{\progchall}{Challenge\xspace}
\newcommand{\progchalls}{Challenges\xspace}
\newcommand{\misconception}{misconception library\xspace}

\newcommand\boxsizecell{6mm}
\newcommand\boxsizefile{5mm}

\tikzset{
  fileframe/.style={rectangle, draw=black, thick,outer sep=0pt,node distance=0pt, minimum size=\boxsizefile},
  filehead/.style={rectangle, draw=black, thick,
  outer sep=-\boxsizefile,node distance=0pt, minimum height=2*\boxsizefile, minimum width=\boxsizefile}
}

\tikzset
  {cell/.style=
    {inner sep=0pt,minimum width=\boxsizecell,minimum height=\boxsizecell,
     fill=white},
   ptr/.style={Circle-stealth,shorten <=-1.5pt},
   void/.style=
    {minimum width=\boxsizecell,minimum height=\boxsizecell,
    fill=white}
  }

\newcommand\ptr[3][]%
  {\node[draw,cell,#1] (#2) {};
   \ifthenelse{\isempty{#3}}%
     {\draw (#2.south west)--(#2.north east);}%
     {\draw[ptr] (#2.center)--(#3);}
  }

\newcommand\ptrEnd[3][]%
  {\node[draw,cell,#1] at (0.5,0.5) (#2) {};
   \ifthenelse{\isempty{#3}}%
     {\draw (#2.south west)--(#2.north east);}%
     {\draw[ptr] (#2.center)--(0.5,0.3);}
  }

\newcommand\element[4][]%
  {\node[draw,cell,#1] (#2) {#3};
   \node[draw,cell,xshift=\boxsizecell] (#2ptr) at (#2) {};
   \ifthenelse{\isempty{#4}}%
     {\draw (#2ptr.south west)--(#2ptr.north east);}%
     {\draw[ptr] (#2ptr.center)--(#4);}
  }

\newcommand\eldots[4][]%
  {\node[draw=none,#1] (#2) {#3};
   \node[draw=none,xshift=\boxsizecell] (#2ptr) at (#2) {};
   \ifthenelse{\isempty{#4}}%
     {\draw (#2ptr.south west)--(#2ptr.north east);}%
     {\draw[ptr] (#2ptr.center)--(#4);}
  }

\begin{document}

\title{Training Students' Abstraction Skills Around a \cafe} % \ed{GB : should we add "in a CS1 course context" ?}
%Learning How to Program Around a \cafe

\author{
  \IEEEauthorblockN{
    G\'eraldine Brieven\orcidlink{0000-0003-1410-1470},Lev Malcev\orcidlink{0000-0002-5259-4336},Benoit Donnet\orcidlink{0000-0002-0651-3398}
  }

  \IEEEauthorblockA{
    Universit\'e de Li\`ege, Institut Montefiore, Belgium
  }
}

\maketitle

% !TEX root = ./paper.tex
\begin{abstract}

Shaping first year students’ mind to help them master abstraction skills is as crucial as it is challenging. Although abstraction is a key competence in problem-solving (in particular in STEM disciplines), students are often found to rush that process because they find it hard and do not get any direct outcome out of it. They prefer to invest their efforts directly in a concrete ground, rather than using abstraction to create a solution.

To overcome that situation, in the context of our CS1 course, we implemented a tool called \cafe. It allows students to actively and regularly practice (thanks to a longitudinal activity) their abstraction skills through a graphical programming methodology.  Moreover, further than reviewing students’ final implementation, \cafe produces a personalized \fb on how students modeled their solution, and on how consistent it is with their final code.  This paper describes \cafe in a general setting and also provides a concrete example in our CS1 course context.  This paper also assesses students' interaction with \cafe through perception and participation data.  Finally, we explain how \cafe could extended in another context than a CS1 course.

%This paper illustrates \cafe in a general setting and also provides a concrete example in a disciplinary-based context, which is our CS1 course.
%CAFE 2.0 allows students to actively and regularly practice their abstraction skills through the \glibpFull methodology. Moreover, further than reviewing students’ final implementation, CAFE 2.0 produces personalized \fb on how students modeled their solution, and on how consistent it is with their final code. Finally, \cafe’s features are embodied in an online activity to make students interact with the platform. That interaction is briefly qualified through students’ actual perception and participation. 
\end{abstract}
%, through two examples of activites running on \cafe.


% !TEX root = ./paper.tex
\section{Introduction}\label{intro}
%%%%%%%%%%%%%%%%%%%%%%%%%%%%%%%%%%%
% ---- WHY
For a first year student, entering Higher Education is a completely other world opening to them compared to Secondary School. First-year students are expected to digest the topics they are taught on their own, with very limited guidance from the supervisors. On one hand, students should become autonomous as it is part of the abilities they should develop and, on the other hand, supervisors cannot often dedicate enough time to each student individually, since they belong to a large and heterogenous group. Moreover, the subjects students have to integrate are often larger and more complex, making the transition even steeper from Secondary School to Higher Education. This situation contributes to a high failure rate, as well as a high withdrawal ratio throughout the year~\cite{tinto}, in particular in CS1 classes~\cite{attrition1,attrition2,attrition3}. This is especially true in our country, where open access to Higher Education is the rule (with some exceptions in the Medicine and Engineering Faculties). The consequence is that we cannot make any kind of assumptions about a first year student's background. This can be a significant drawback in some areas, like Computer Science~\cite{math_progra}, that strongly rely on Mathematical skills. Shaky foundations in Mathematics leads to poor abstraction capacities as well as a lack of rigor in problem solving, while abstraction and problem solving represent the core of the Computer Science curriculum. However, many students are not fully aware of that until they take the first evaluation, which reveals where they stand with respect to the curriculum requirements. In many cases, that check point already comes too late in the semester : students feel demotivated and cannot make up the time they have lost as new topics keep being taught.

\begin{figure}[!t]
  \begin{center}
      \includegraphics[width=8cm]{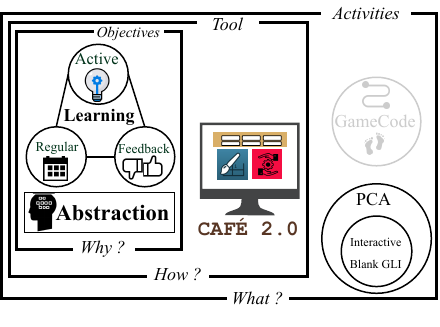}
  \end{center}
  \caption{Motivation and context around \cafe and our CS1 course.}
  \label{cafe.whyHowWhat.fig}
\end{figure}

In that teaching context, one of our goals in our Introduction to Programming class (usually refered to as ``CS1'') is to maintain students onboard. To do so, regular activities are organised~\cite{info0946_afl} to give them the opportunity to be actively learning and to receive \fb. Most of their productions mobilize abstraction skills since any Computer Scientist should demonstrate such abilities~\cite{abstraction_importance, abstraction_prereq}. More generally, abstraction skills are involved in all STEM (Science, Technology, Engineering, and  Mathematics) disciplines because they support problem and solution modeling, whatever the nature of the problem~\cite{abstraction_stem}. Thanks to abstraction, a whole class of equivalent problems (where only input parameters vary) can be addressed, rather than only some specific instances. In our course, students apply abstraction through the \glibpFull (\glibp)~\cite{glibp}, consisting in representating and manipulating a drawing reflecting a solution supported by an iterative process (i.e., a loop in a piece of code). That context and goals are illustrated in the rectangle labeled ``Objectives'' in Fig.~\ref{cafe.whyHowWhat.fig}.

% ---- HOW : CAFÉ
Encompassing that rectangle, Fig.~\ref{cafe.whyHowWhat.fig} expresses how abstraction can be taught with respect to those objectives. In practice, the only way to make a large group of students active and provide them with a correct and personalized \fb is to transit through a remote activity supported by an online system. We implemented such a learning tool called \cafe (standing for ``Correction Automatique et Feedback des \'{E}tudiants''). It is currently applied only in the context of our CS1 class in which students are exposed to C programming language concepts and a graphical programming methodology (\glibp). \cafe automatically assesses students' programming exercises and provides students with high quality \fb and \ffw information (i.e., what should they do to improve their solution). One key point of \cafe is that it does not only focus on the program output but also on the cognitive abstraction process inherent to the program construction, which differs from a large majority of learning tools restricting to automatic code simulation and assessment~\cite{webcat, problets, inginious, codingbat, myproglab, unlock,coderunner,tool_dodona, tool_vide}. That differentiation gives to \cafe the potential to be transposed in other (STEM) disciplines that would rely on a sequential resolution process founded on a scheme constructed upstream, similarly to our course.

% ---- WHAT : ACTIVITIES
To make students use a learning tool, it needs to be integrated into the course via activities, as shown in the outer rectangle labeled ``Activities'' in Fig.~\ref{cafe.whyHowWhat.fig}. In our course, the \dfn{Programming Challenge Activity} (\pca)~\cite{pca} was created. It spans the four-month semester by regularly addressing statements students should solve by submitting some solutions as many times as they want. Furthermore, for each submission, students receive personalized \fb (especially detailed \fb about their solution modeling through a drawing) and \ffw. In addition to that activity, it is aimed to offer students another kind of opportunity to train abstraction skills. To meet that purpose, the \gamecodes~\cite{gamecode} are currently being implemented. Contrary to the \pca, the \gamecodes give students more freedom and guidance across their resolution.

In this paper, we carefully depict \cafe features, by defining them from a generic and specific perspective.  Doing so, we highlight the interest and potential of \cafe out the scope of our CS1 class.  Then, this paper discusses two remote activities : the \pca (implemented over \cafe) and the \gamecodes (under implementation). After that, we report how students receive the \pca and \cafe during our course organized during Academic Year 2022--2023. Finally, this paper exposes how \cafe could be extended to support new problem profiles (from other STEM fields).  In particular, we define the checklist a resolution flow should follow to get the best from \cafe.

The remainder of this paper is organized as follows: Sec.~\ref{cafe} depicts \cafe, while Sec.~\ref{activity} discusses activities in \cafe; Sec.~\ref{eval} presents perception and participation data we collected; Sec.~\ref{extension} explains how \cafe should be extended to support additional disciplines; Sec.~\ref{related} positions this paper with respect to the state of the art; finally, Sec.~\ref{ccl} concludes this paper by summarizing its main achievements.

\section{\cafe}\label{cafe}
%%%%%%%%%%%%%%%%%%%%%%%%%%

%\subsection{Origin of \cafe}\label{cafe.10} %: \cafeun
%%%%%%%%%%%%%%%%%%%%%%%%%%%%%%%%%%%%%

\begin{figure}[!t]
  \begin{center}
      \includegraphics[width=8cm]{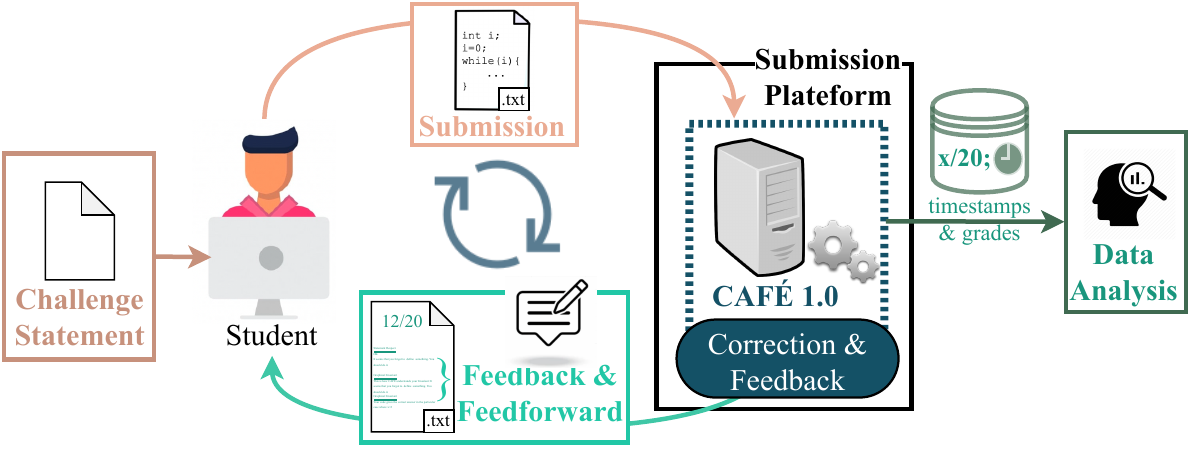}
  \end{center}
  \caption{Illustration on how students interact with \cafeun.}
  \label{cafe.10.fig}
\end{figure}

\cafe has been implemented as a new version of \cafeun. Initially, \cafeun emerged as a set of Python scripts integrated in a submission plateform~\cite{cafe}.  As shown in Fig.~\ref{cafe.10.fig}, to interact with \cafeun, the students need to solve some statement, format their solution in a text file with predefined placeholders, and then upload it on a submission plateform. Each submission is instantaneously processed by \cafeun that computes the grades, highlights what should be adapted in the current submission (through the \fb), and provides pointers to the theoretical courses (through the \ffw). In this way, students get the opportunity to realize their misunderstanding and improve their subsequent submissions. Fig.~\ref{cafe.10.fig} also highlights that, as supervisors, in addition to be timesaving and scalable, such a system allows us to keep track of student’s behavior by collecting basic data related to their activity and performance.

However, in practice, that initial version has several drawbacks, i.e., ($i$) the absence of interactiveness (since \cafeun has no interface), ($ii$) the lack of guidance across the resolution (since it has to be handled beforehand)~\cite{lessonLearned}, and, ($iii$), the limited learning analytics collected. That prevents a student from tracking their progress and learning journey. Likely, those limitations were repeling some students from taking the full benefits from that learning approach. That is what leveraged the development of \cafe, turning \cafeun into a fully online plateform and expanding it in order to make students' online-learning journey richer and more natural. The different upgrades from \cafeun are detailed in Sec.~\ref{cafe.20}.

\subsection{\cafe Overview}\label{cafe.20}
%%%%%%%%%%%%%%%%%%%%%%%%%%%

In this subsection, \cafe is presented from two perspectives : a generic one (providing a high level overveiw of \cafe, so that the reader can project it more easily in their own discipline ground) and a specific one (allowing to embody that generic view).

\begin{figure*}[!t]
  \begin{center}
    \subfloat[Generic modules]{
      \label{cafe.overview.fig.generic}
      \includegraphics[width=0.8\columnwidth]{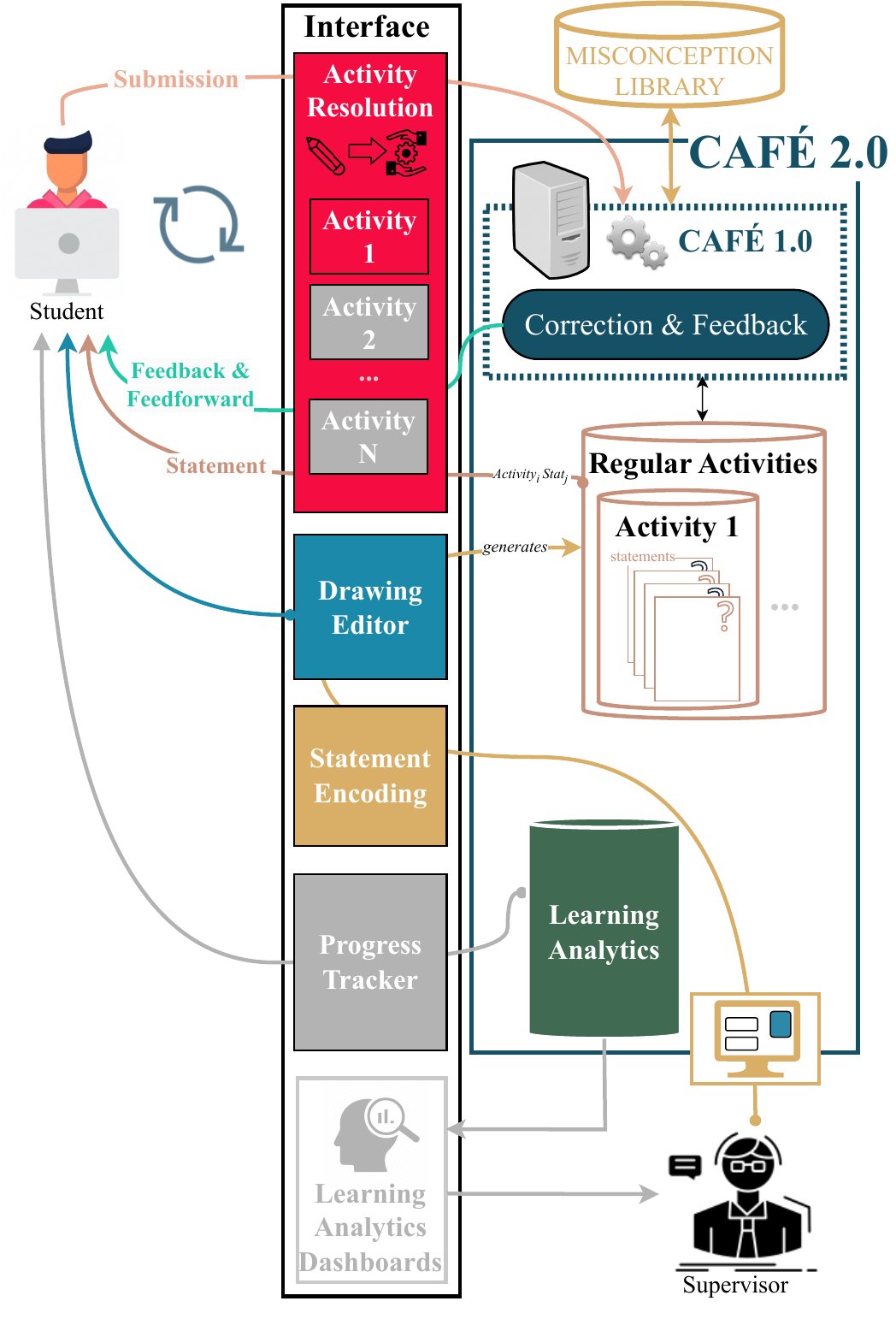}
    }
    \subfloat[Modules as implemented in our CS1 course.]{
      \label{cafe.overview.fig.cs1}
      \includegraphics[width=1.1\columnwidth]{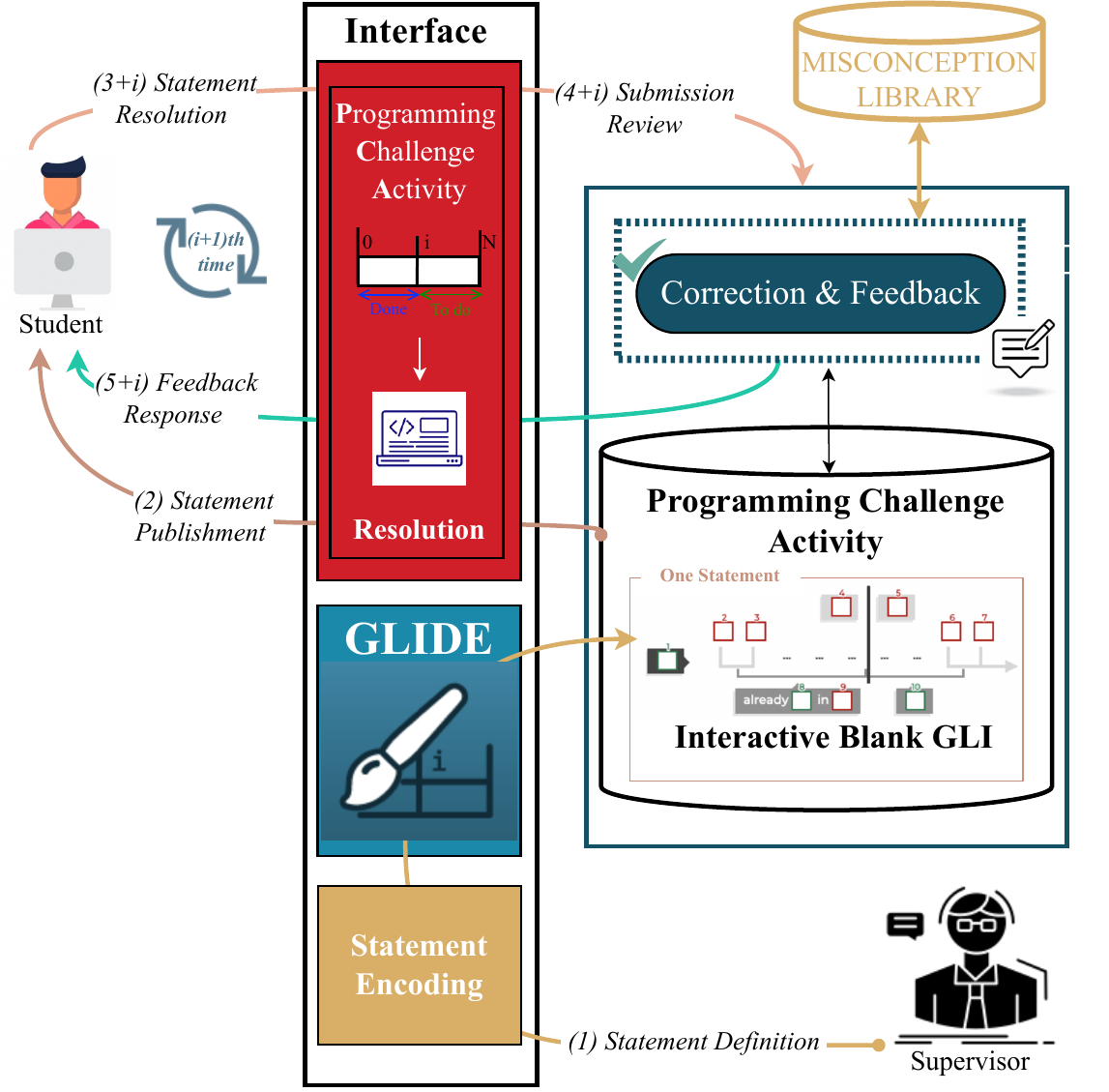}
    }
  \end{center}
  \caption{High level overview of \cafe infrastructure.}
  \label{cafe.overview.fig}
\end{figure*}

From a generic point of view, Fig.~\ref{cafe.overview.fig.generic} illustrates the different functionalities \cafe is currently offering. Those functionalities target two types of end users: the students and the supervisors. Considering the students, they can interact with three different modules : ($i$) the \dfn{Activity Resolution}, where students can pick some statement, solve it, and submit it in order to receive a personalized \fb and \ffw ; ($ii$) the \dfn{Drawing Editor}, equipping students with graphical components they can drag and drop in order to design some solution according to some outline being shown and detailed during the course ; ($iii$) the \dfn{Progress Tracker}, whose goal is to depict where students stand in their learning journey, with respect to their activity and performance on the tool. Besides this, a supervisor interacts with three modules that echo the ones intended for the students : ($i$) the \dfn{Statement Encoding}, through which a supervisor can define a new statement and parametrize its automatic assessment, \fb, and \ffw, in the context of an activity ; ($ii$) the \dfn{Drawing Editor}, that may be used to define a new statement, if the supervisor wants to include some schemes to be completed ; ($iii$) the \dfn{Learning Analytics Dashboard}, illustrating the students' learning behavior, based on the collected learning analytics. Such a feature will automatically distill students' performance thanks to the personalized \fb that is constructed, which also allows us to identify precisely the points of misunderstood material.

In Fig.~\ref{cafe.overview.fig.generic}, the modules that are usable via an interface (i.e., frontend) are comprised in the central rectangle (``Interface''). Then, the backend is represented on the right where the main data that needs to be stored to support those modules is represented through cylinders. Finally, the backend also includes the \textit{Correction and \Fb} functionality responsible for handling a student's submission, knowing to which statement it is supposed to respond to and relying on a \misconception where typical mistakes made by students have been stored.
%In practice, the bottom functionalities (\textit{Progress Tracker} and \textit{Learning Analytics Dashboard}) are still under implementation, that's why there won't be any dedicated focus on them.

Next, from a particular point of view, Fig.~\ref{cafe.overview.fig.cs1} refreshes Fig.~\ref{cafe.overview.fig.generic} by specifying the different modules that have just been exposed in the context of our CS1 course. More precisely, Fig.~\ref{cafe.overview.fig.cs1} focuses on the modules that are currently implemented (colored in Fig.~\ref{cafe.overview.fig.generic}) and numbers the flow that must be followed by a statement, from its definition to its resolution (that may rely on several improvement iterations). Considering the different modules, it can be noticed that the generic Drawing Editor defined previously has been instantiated as \glide. It provides pre-defined patterns and tutorials, specific to the programming methodology being taught in our course~\cite{glibp}. Next, the activity that is proposed as a concrete opportunity to practice is the \dfn{Programming Challenge Activity} (\pca)~\cite{pca}. It mainly consists in addressing some statements to the students whose expected resolution namely relies on an interactive blank outline (referred by \bgli) to fill in, as shown on the right of Fig.~\ref{cafe.overview.fig.cs1}. All those modules implemented in the context of our CS1 course are detailed in the next subsections.

%\subsection{Backend of \cafe}\label{cafe.backend}
%%%%%%%%%%%%%%%%%%%%%
%\ed{BD: This section aims at describing \cafe backend, in particular the various modules: Correction (Sec.~\ref{cafe.backend.correction}), Activity (Sec.~\ref{cafe.backend.activity}), Graphical Reasoning (Sec.~\ref{cafe.backend.glide}), and Learning Analytics (Sec.~\ref{cafe.backend.la}).  Again, we need a figure illustrating the relationship(s)/interaction(s) between the various modules.

\subsection{Activity Resolution Module}\label{cafe.activity}
%%%%%%%%%%%%%%%%%%%%%%%%%%%%%%%%%%%%%%%%

\begin{figure*}[!t]
    \subfloat[Generic resolution flow through productions.]{\label{cafe.actGen.resol.fig}
      \includegraphics[width=9cm]{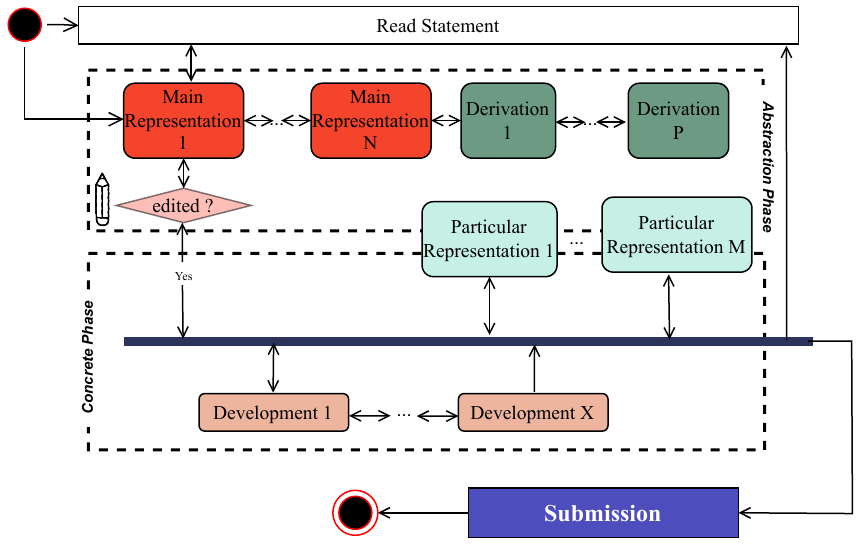}
    }
    \subfloat[Resolution flow relying on the \gli.]{\label{cafe.actSpec.resol.fig}
      \includegraphics[width=9cm]{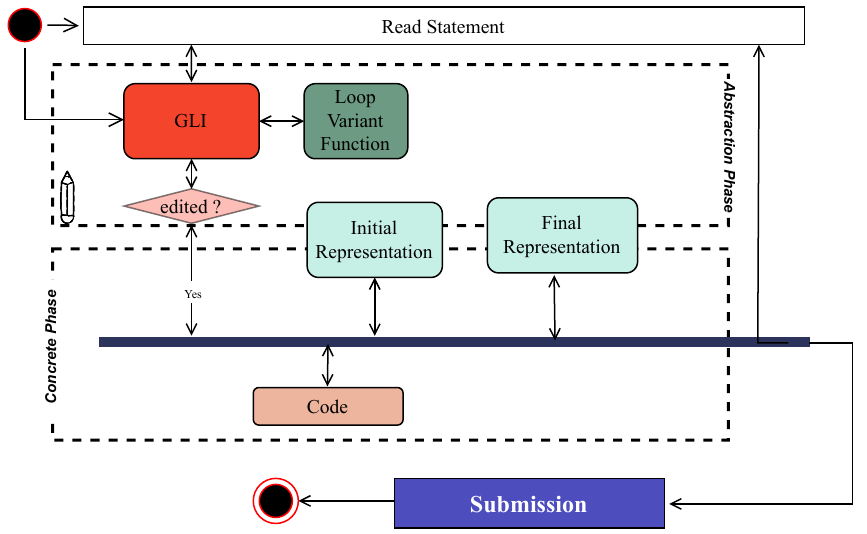}
    }
  \caption{Resolution module of \cafe.}
  \label{cafe.resolutionAct.fig}
\end{figure*}

\begin{figure*}[!t]
  \begin{center}
    \includegraphics[width=16cm]{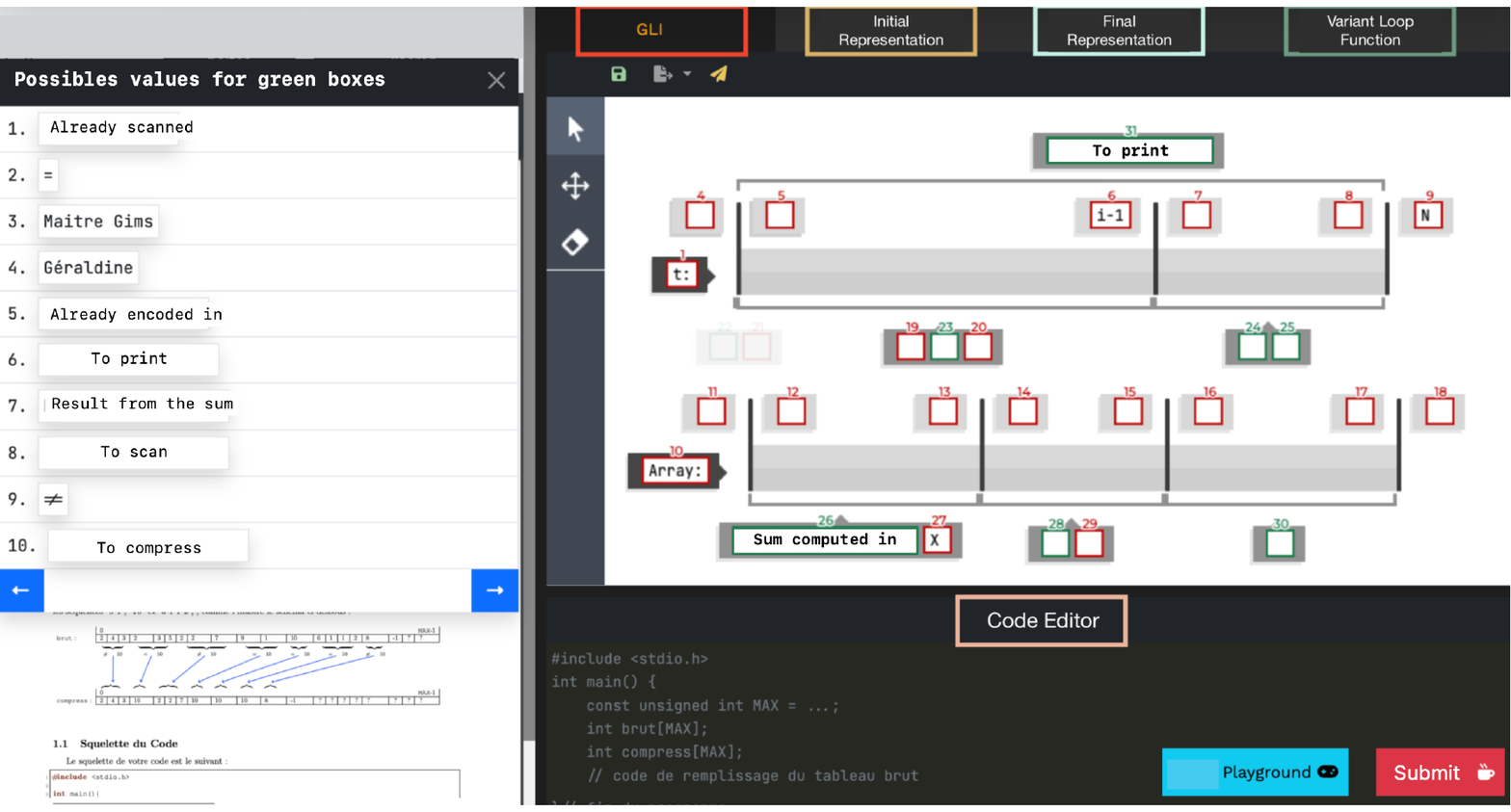}
  \end{center}
\vspace{-1cm}  
  \caption{\bgli in the \pca. It also shows how our tool follows the \glibp methodology with tabs, one for each step of the resolution process.}
  \label{cafe.frontend.fig.bgli}
\end{figure*}

Similarly to \cafe's overview, the \textit{Activity Resolution} is presented at two levels : a generic one and a specific one. First, Fig.~\ref{cafe.actGen.resol.fig} shows that any resolution supported by \cafe should be composed of an \absphase followed by a \devphase. Then, for each phase, those can include one or more sequential productions~\cite{glibp_abs}, one production being illustrated in a rounded rectangle. Additionally, some productions may also overlap those two phases in order to bridge them based on specific configurations of their solution. Finally, some locks can be defined between the productions, as represented through the diamond ``edited''. In Fig.~\ref{cafe.actGen.resol.fig}, it comes just after the ``Main Representation'', meaning that students need to first work on it before deriving some specific states and developing their solution. In regards to that, Fig.~\ref{cafe.actSpec.resol.fig} depicts that resolution process in the context of our CS1 course. It results in five productions in our case. More concretely, those are also shown through Fig~\ref{cafe.frontend.fig.bgli}. On that figure, the four tabs above support the abstraction of the solution via the \bgli (described in Sec.~\ref{cafe.prodModelisation}) and its transposition into concrete states thanks to movable bars. In particular : ($i$) ``GLI'' consists in filling the \bgli, turning it as students' own \gli; ($ii$) ``Initial Representation'' requires students to graphically manipulate their \gli to reflect the initial configuration of their solution. In this way, they can derive how the variables supporting their solution should be declared and initialized in their code; ($iii$) ``Final Representation'' corresponds to graphical manipulation of the \gli to illustrate the final solution so that students can deduce under which condition their loop stops; ($iv$) ``\lv'' is for proposing a function that gives the number of elements that still need to be processed in order to get the final solution. All those tabs are supposed to be done in sequence. Students have also access to the code editor (bottom part of Fig.~\ref{cafe.frontend.fig.bgli}, labelled as ``Code Editor'').  The code may be pre-filled with a template~\cite{automated_feedback_survey} students must edit with their code. Students have also access to the ``playground'' mode in which they can compile and test their pieces of code. Once students are ready, they can submit their whole solution.

At that point, two interests in decomposing students resolution into pre-defined productions can be highlighted. First, it allows \cafe to pave students' resolution with respect to a given methodology. Next, it frames their solution, making feasible an automatic correction and personalized \fb mainly thanks to the \bgli.

\subsection{Correction and \Fb Module}\label{cafe.correction}
%%%%%%%%%%%%%%%%%%%%%%%%%%%%%%%%%%%%%%%%%%%

\begin{figure*}[!t]
    \subfloat[Generic \fb module.]{\label{cafe.actGen.fb.fig}
      \includegraphics[width=1\columnwidth]{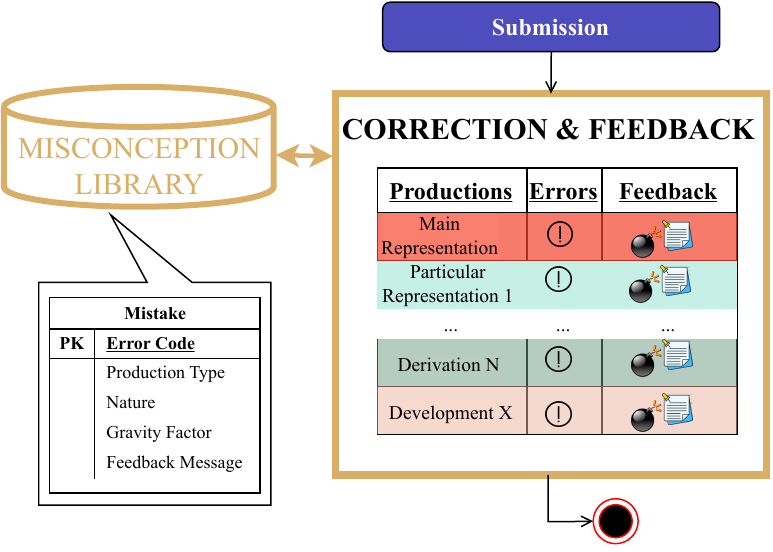}
    }
    \subfloat[Example of \fb based on a given \misconception.]{\label{cafe.actSpec.fb.fig}
      \includegraphics[width=1\columnwidth]{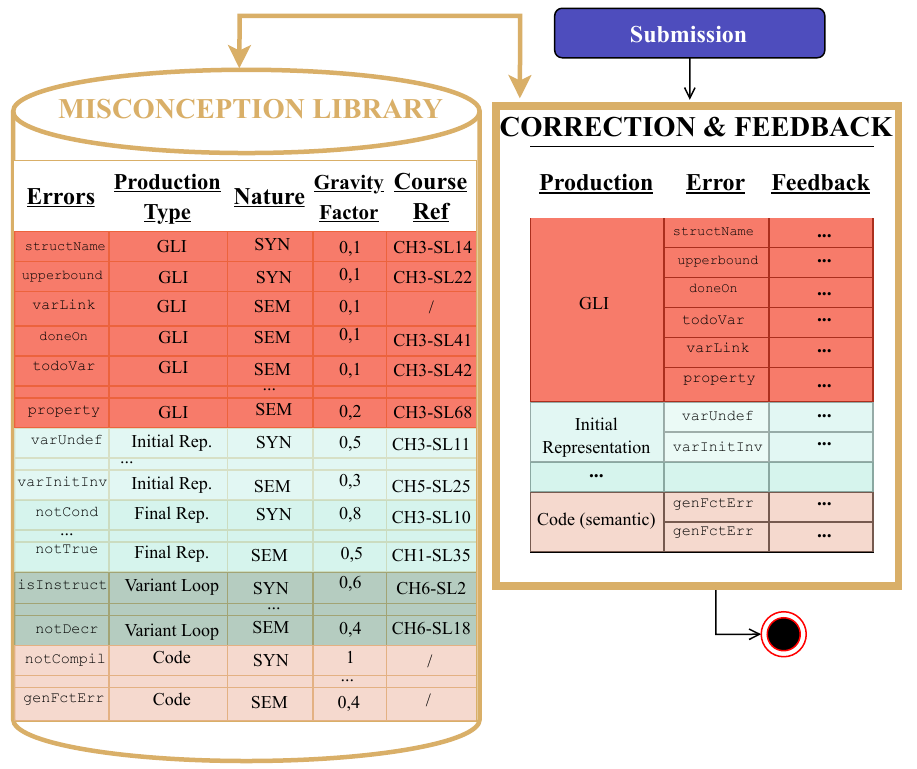}
    }
  \caption{Correction and \Fb module of \cafe.}
  \label{cafe.FB.fig}
\end{figure*}

Fig.~\ref{cafe.FB.fig} presents how a given student's solution gets instantaneously assessed and commented, based on a predefined \misconception. Like previously, that process is represented in a general way (through Fig.~\ref{cafe.actGen.fb.fig}) while Fig.~\ref{cafe.actSpec.fb.fig} instantiates it in the context of our course. Both figures show that the correction and \fb are performed based on a \misconception containing typical mistakes students tend to make. It is worth noticing that the misconception concept is broadly used in the STEM literature\cite{misconc_usage,misconc_usage1}. In this paper, we similarly use the terms ``misconception'', ``error'', and ``mistake'' to point out ``something that is done wrong''.

\subsubsection{The \misconception}\label{cafe.fb.misconLib}
%%%%%%%%%%%%%%%%%%%%%%%%%%%%%%%%%%%
From a general point of view, any supervisor wanting to use \cafe as an automatic assessment and \fb system should define a rubric checklist~\cite{rubricDef} beforehand, forming so the \dfn{\misconception}. That rubric should be organised according to the productions a student's submission is made up of. For each production, typical mistakes should be identified (based on previous experiences, like presented in other studies~\cite{misconc_identif}). Then, each mistake should be characterized by a unique error code, a nature (syntactic/semantic), a gravity factor (quantifying how serious the mistake is), a \fb message (explaining in details the error), and, optionally, a corresponding reference to the course (i.e., \ffw). Once the \misconception has been fed, some respective rule-based checks must be implemented and simply configured in order to catch each mistake based on a given submission.

\subsubsection{The Correction and \Fb Construction}\label{cafe.fb.constr}
%%%%%%%%%%%%%%%%%%%%%%%%%%%%%%%%%%%%%%%%%%%%%%%%%%%%
When those last set-ups are ready, the system can process a given submission. For a given student's submission, each production is digested by a dedicated checker module that detects any potential mistake defined in the \misconception. If a mistake is captured, the student's final grade gets impacted with respect to the gravity factor characterising the mistake. In addition to this, the corresponding \fb message and reference to the course are added in the list of comments being provided to the student, eventually. That list of comments is splited on a per production basis.

Fig.~\ref{cafe.actSpec.fb.fig} shows an example of \misconception from which mistakes were detected in a given submission. Once the \fb has been received, a student may improve their solution and submit it again.

\subsection{Production Modeling}\label{cafe.prodModelisation}
\begin{figure}[!t]
  \begin{center}
    \includegraphics[width=7cm]{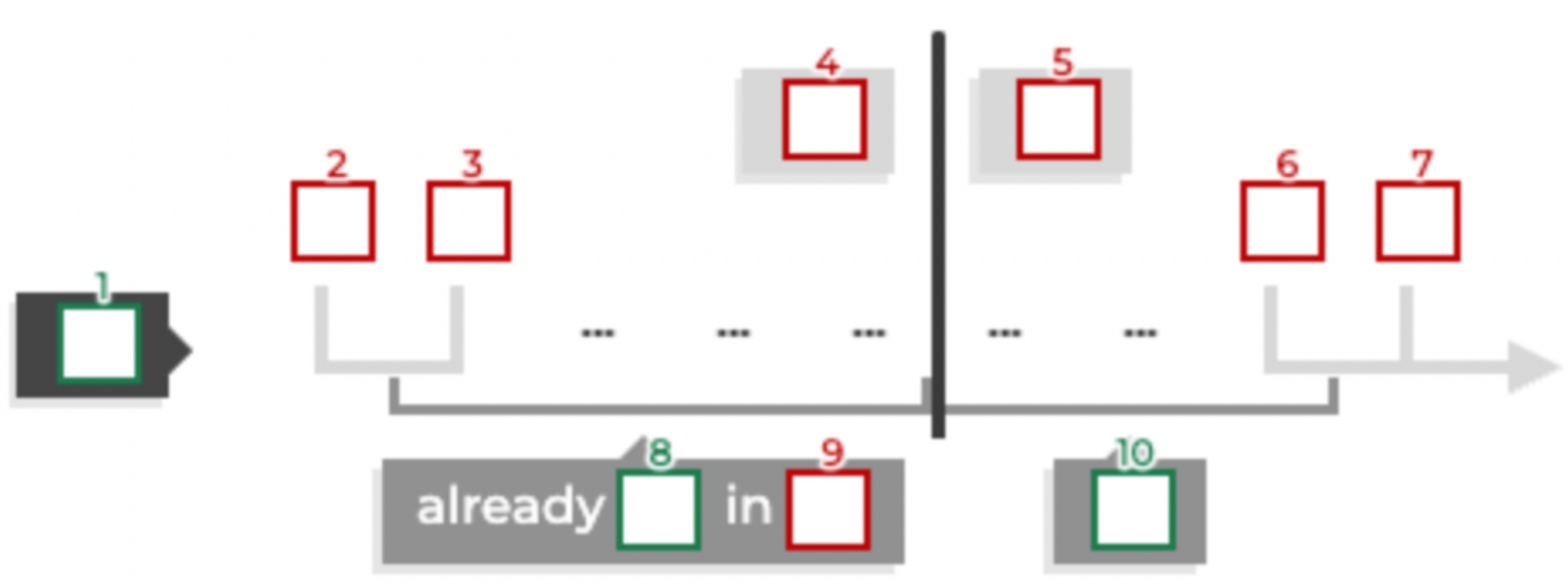}
  \end{center}
  \caption{\bgli as solution ground.}
  \label{cafe.bgli.fig}
\end{figure}

%%%%%%%%%%%%%%%%%%%%%%%%%%%%%%%%%
For a given production (a production being represented in a rounded rectangle in Fig.~\ref{cafe.resolutionAct.fig}), in order to be able to capture errors, a tradeoff must be found between constraining the solution and letting enough freedom to students. The more bounded the solution, the more predictable the students' answers with respect to the typical mistakes, which can be caught through rule-based checks. Besides this, The more freedom students get, the easier the transposition of their own reasoning to the provided canvas. In practice, one way is to shape each production and model the expected solution using blank solution components whose semantic and relations between each others must be specified beforehand. Typically, in our course, solution components stand as fields to fill, instructions in the code, or components of the \bgli (being movable bars and boxes). In particular, the \bgli is a blank drawing depicting only the general shape that should follow a correct and rigorous \gli~\cite{glibp,glibp_abs}. Students must then annotate properly the figure so that the drawing becomes their \gli modeling their own solution. An example of \bgli is provided in Fig.~\ref{cafe.bgli.fig}. Any \bgli always comes with two types of boxes: ($i$) red boxes standing to host expressions (i.e., constants, variables, operations, or left blank) and are to be completed by students without support; ($ii$) green boxes standing to host labels that students must drag and drop from a pre-defined list (see the list on the left of the \bgli in Fig.~\ref{cafe.frontend.fig.bgli}).

That list contains multiple choices, some of them being the expected answers, others being purely random.  Doing so, we pave the way for an automatic correction of the \gli (with strong \fb and \ffw). This can be achieved thanks to the fact each box is numbered.  In this way, when a student's solution gets corrected, each piece of the solution is easily pointed out, allowing to bring a rich \fb while still keeping it clear and smooth to digest for the student. To define a \bgli, a supervisor can use the drawing editor \glide.

%In this way, consistency checks can be performed inside a production and between productions.
%Like spliting the resolution into pre-defined productions, defining a \bgli is purposeful at two aspects. First, it provides students with a structured and coherent framework so that they do not start their loop design from scratch. To meet that purpose, as early stage, the \bgli method is proposed. Then, by framing their solutions through a given canvas, the semantic of a students' solution can be automatically corrected and commented. As most of the activities in our course consist in writing loops and as the course requires to write loops based on the \gli, any programming activity must embed \invariants so that students can train themselves. At first, it may appear difficult to combine automatic correction and graphical representation. We solve this by asking students to fill in a \bgli.

\subsection{Drawing Editor Module}\label{cafe.glide} % Graphical Reasoning
%%%%%%%%%%%%%%%%%%%%%%%%%%%%%%%%%%%
\begin{figure}[!t]
  \begin{center}
      \includegraphics[width=8cm]{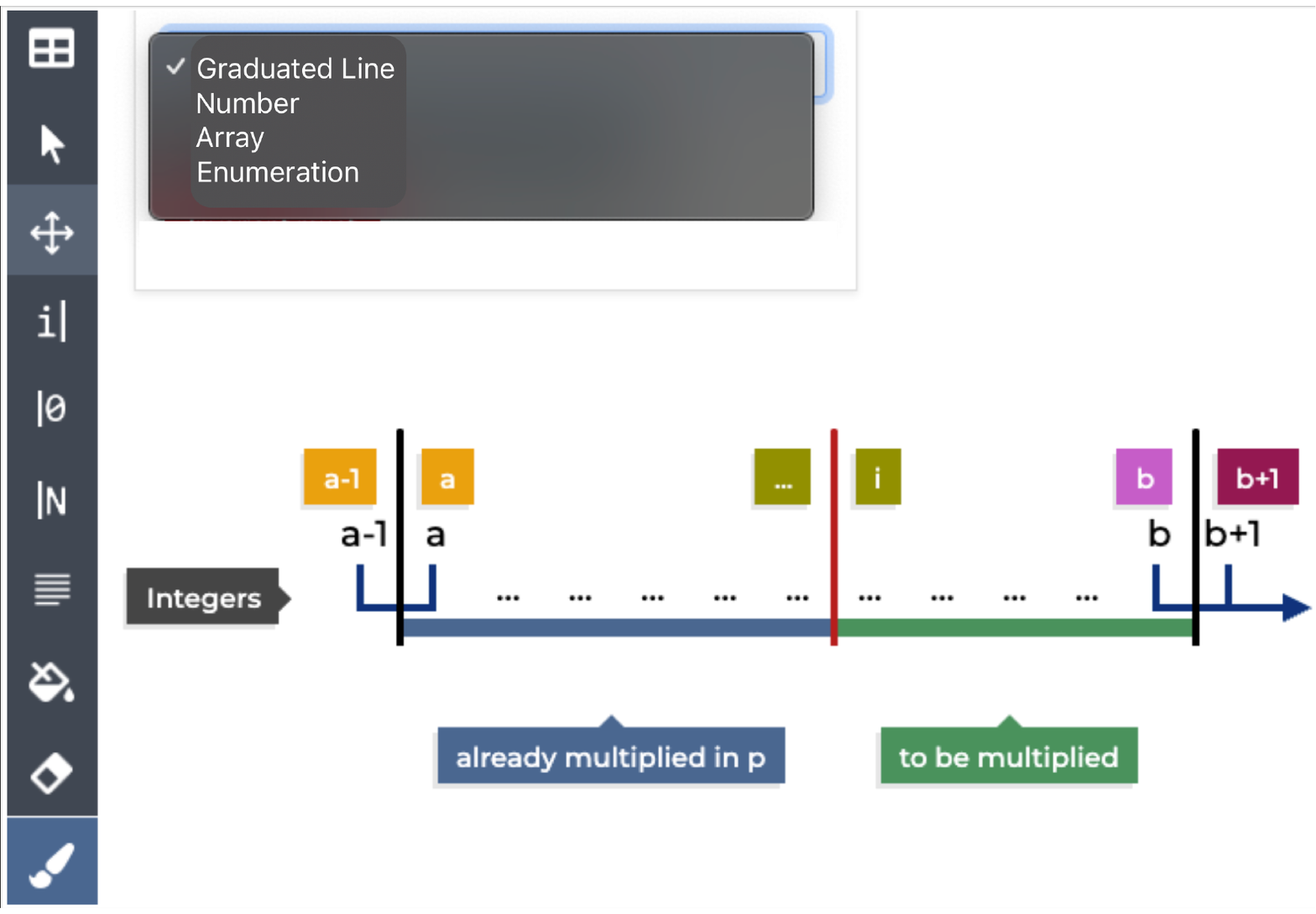}
  \end{center}
  \caption{Screenshot of \glide.}
  \label{glide.fig}
\end{figure}

The \gli Drawing Editor (\glide) proposes to supervisors all the components a \gli can be composed of so that they can build up some \bgli, responding to a given problem.

Besides this, \glide also helps students in drawing their own \gli by proposing pre-defined graphical components~\cite{glibp} students must arrange and fill in. Fig.~\ref{glide.fig} shows a final representation of a \gli, illustrating how to compute the product of all integers belonging to a range specified as input. Furthermore, students can be guided across their composition by activating some step-by-step tutorials. Once a student considers their \gli is completed, they can submit it and some basic checks are performed. In particular, syntactic mistakes are detected (such as the lowerbound being further than the upperbound or some description of what has been achieved so far that is missing). However, the \gli semantic is not verified, which means that the solution can be positively assessed by the \glide although the \gli does not make sense.

Similarly to that particular instance, a Drawing Editor could be implemented in other fields by equipping it with the adequate graphical components, modeling them in order to define rules on and drawing up a step-by-step tutorial guiding students in handling those components.

%\subsection{Learning Analytics Module}\label{cafe.la}
%%%%%%%%%%%%%%%%%%%%%%%%%%%%%%%%%%%%%%%%%%%%%%%%%%%%%

%\subsection{Frontend of \cafe}\label{cafe.frontend}
%%%%%%%%%%%%%%%%%%%%%

%For students, connecting on \cafe is achieved through our University single-sign out (SSO) process.  Once connected, \cafe loads the welcome page, illustrated in Fig.~\ref{cafe.frontend.fig.welcome}.  Four possibilities are offered to students: ($i$) using \glide, our \gli Drawing Editor~\cite{glibp}, ($ii$) realising a Challenge~\cite{pca}, the remote programming activity automatically corrected by \cafe, ($iii$) doing a \gamecode~\cite{gamecode}, a remotely supervised programming exercise in which students follows their own resolution path, and, ($iv$) a progress tracker that shows to students where there are with respect to the course evolution.\footnote{As suggested by Fig.~\ref{cafe.frontend.welcome}, the \gamecode and the progress tracker are currently under construction.} Playground with submission area is called ``Solution Template'' by Keuning et al.~\cite{automated_feedback_survey}

% !TEX root = ./paper.tex
\section{Designing an Activity}\label{activity}
%%%%%%%%%%%%%%%%%%%%%%%%%%%%%%%
Automatic assessment and \fb can be fully beneficial only if it gets encapsulated in some activities with a sound pedagogical reflection behind~\cite{activitySupportAutomaticAss}. In this section, two activities are introduced : the Programming Challenge Activity (\pca)~\cite{pca} and the \gamecodes~\cite{gamecode} . The \pca is already available in \cafe while the \gamecodes are currently under construction.

\subsection{The \pca (relying on the \bgli)}\label{pca}
%%%%%%%%%%%%%%%%%%%%%%%%%%%%%%%%%%%%%%%%%%%%
\begin{figure}[!t]
  \begin{center}
    \includegraphics[width=8cm]{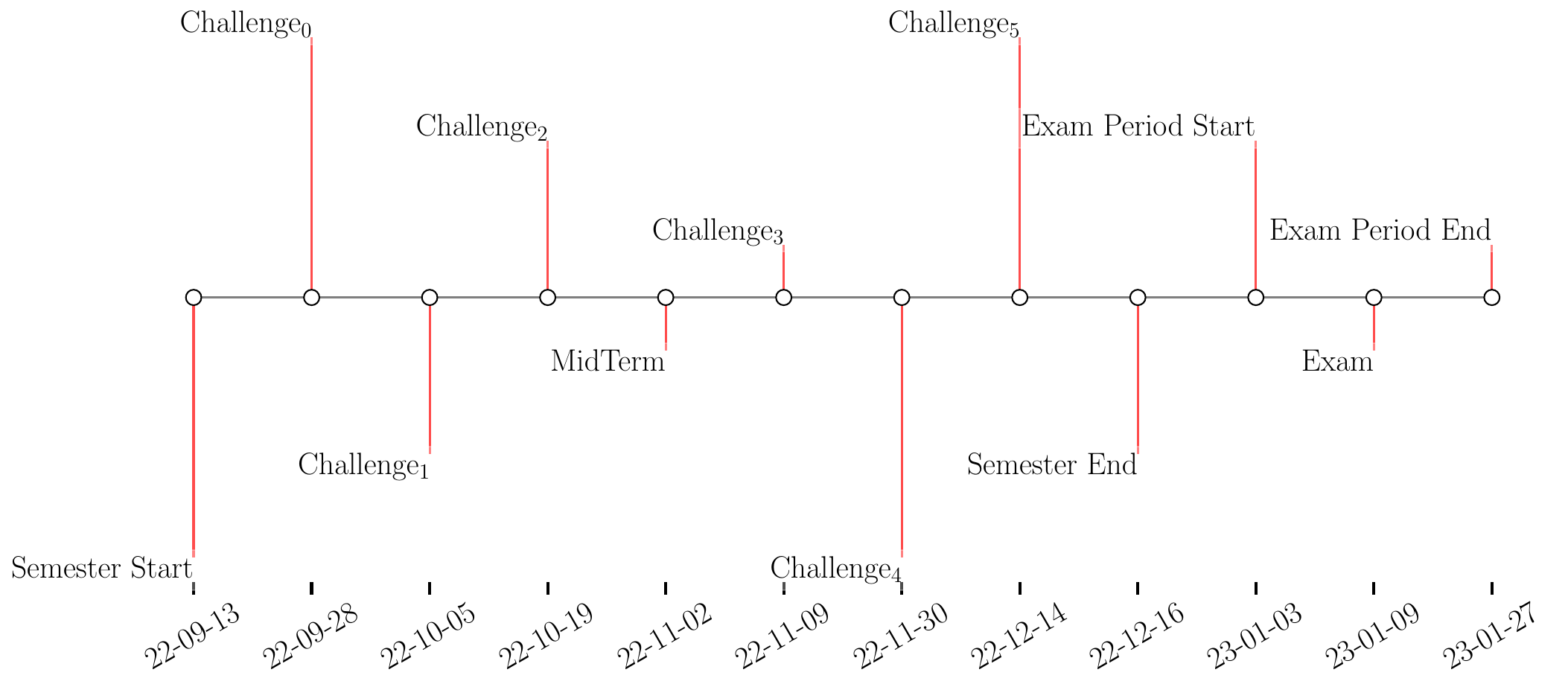}
  \end{center}
  \caption{\cafe activity timeline over the semester.}
  \label{pca.timeline}
\end{figure}

The \pca is made up of six \progchalls. A \progchall is a statement aligned to one or several theoretical chapters taught the week(s) before. In the fashion of the chapters in our CS1 course, the \progchalls are cumulative, requiring a good level of understanding about the previous topics to be properly handled. Each \progchall consists in producing some pieces of code. For \progchalls 2, 3 and 4, students must also graphically model their solution by filling some \bgli~\cite{glibp}. Regarding the modalities, any \progchall starts on Wednesday, 06:00PM and finishes on Friday, 08:00PM. During this 2-days timeframe, a student can submit up to three times their solution, each one receiving an automatic \fb and \ffw.  The latest submission determines the final mark and each \progchall accounts for 2\% of the final mark for the course. After that certificative period, students are free to keep training, but it will not affect their final grades.
It is worth noting that the first \progchall (called Challenge 0) does not account in the final grade. Its only purpose is to make sure students grasped how to use \cafe.  Finally, each student has the opportunity to play a \dfn{trump card} allowing them to skip one of the Challenges (that, will not count towards the student's mark)~\cite{pca}. Fig.~\ref{pca.timeline} illustrates the timeline of the \pca for Academic Year 2022--2023.  The six \progchalls were spread between September, \nth{13} and December, \nth{16}.  The blocus (i.e., a period during which no classes are organized and students are supposed to prepare themselves for the upcoming exams) is organized between December, \nth{16} and January \nth{3} for Academic Year 2022--2023.  Fig.~\ref{pca.timeline} also shows when the midterm and the final exam were organized.

\subsection{The \gamecodes}\label{gamecodes}
%%%%%%%%%%%%%%%%%%%%%%%%%%%%%%%%%%%%%%%%%%%%
Contrary to the \pca, the purpose of the \gamecodes is to give students the opportunity to get a personalised learning experience across exercises resolution~\cite{gamecode}. A \gamecode corresponds to a large statement to solve through predefined resolution steps. In our case, one \gamecode is defined per chapter of the course. Students are fully free to take them or not, as they are not certificative. Furthermore, when they solve a \gamecode, students can choose how to handle it by jumping to the resolution step they want at any time, asking for tips or theoretical reminders and submitting answers to get \fb.

% !TEX root = ./paper.tex
\section{Preliminary Evaluation}\label{eval}
%%%%%%%%%%%%%%%%%%%%
This section discusses some observations from the data collected by \cafe.  In particular, we focus on how students grasp the tool.  Sec.~\ref{eval.dataset} explains the data we collected, while Sec.~\ref{eval.results} focuses on metrics of interest.

\subsection{Dataset}\label{eval.dataset}
%%%%%%%%%%%%%%%%%%%%%
%We deployed \cafe in our CS1 course.  The course is given during the first semester that spans between mid-September until Christmas.  Exams are organized in January.  The period between the end of the semester and the exams is called \dfn{blocus} and is used for studying.
%\cafe was used to support a Programming Challenge Activity (\pca)~\cite{pca} in which students are exposed to six programming Challenges whose modalities are explained in Sec.~\ref{pca}. \ed{BD; I don't think those two paragraphs are still required as they simply repeat things developped in the Activity Section.}

During Academic Year 2022--2023, 97 students ($N_r$=97) registered to our CS1 class.  Among them, 76\% were new comers (first year at University), while 23\% were either repeaters or have changed their programs.

We collected data throughout the semester thanks to \cafe Learning Analytics module.  Among the registered students, a maximum of 80 students ($N_c$=80 -- 82.5\% of $N_r$) connected at least once on \cafe.

A survey was conducted at the end of the final exam. 74 students ($N_s$=74 -- 76.3\% of $N_r$) shared their opinion. The survey included Likert scale and open questions, all related to different aspects of \cafe.

\subsection{Results}\label{eval.results}
%%%%%%%%%%%%%%%%%%%%%%%%%%%%%%%%%%%%%%%%
Results include students’ view regarding \cafe (Sec.~\ref{eval.results.perception}) as well as how they were active on \cafe (Sec.~\ref{eval.results.participation}).

\subsubsection{Perception}\label{eval.results.perception}
\begin{figure}[!t]
  \begin{center}
      \includegraphics[width=8cm]{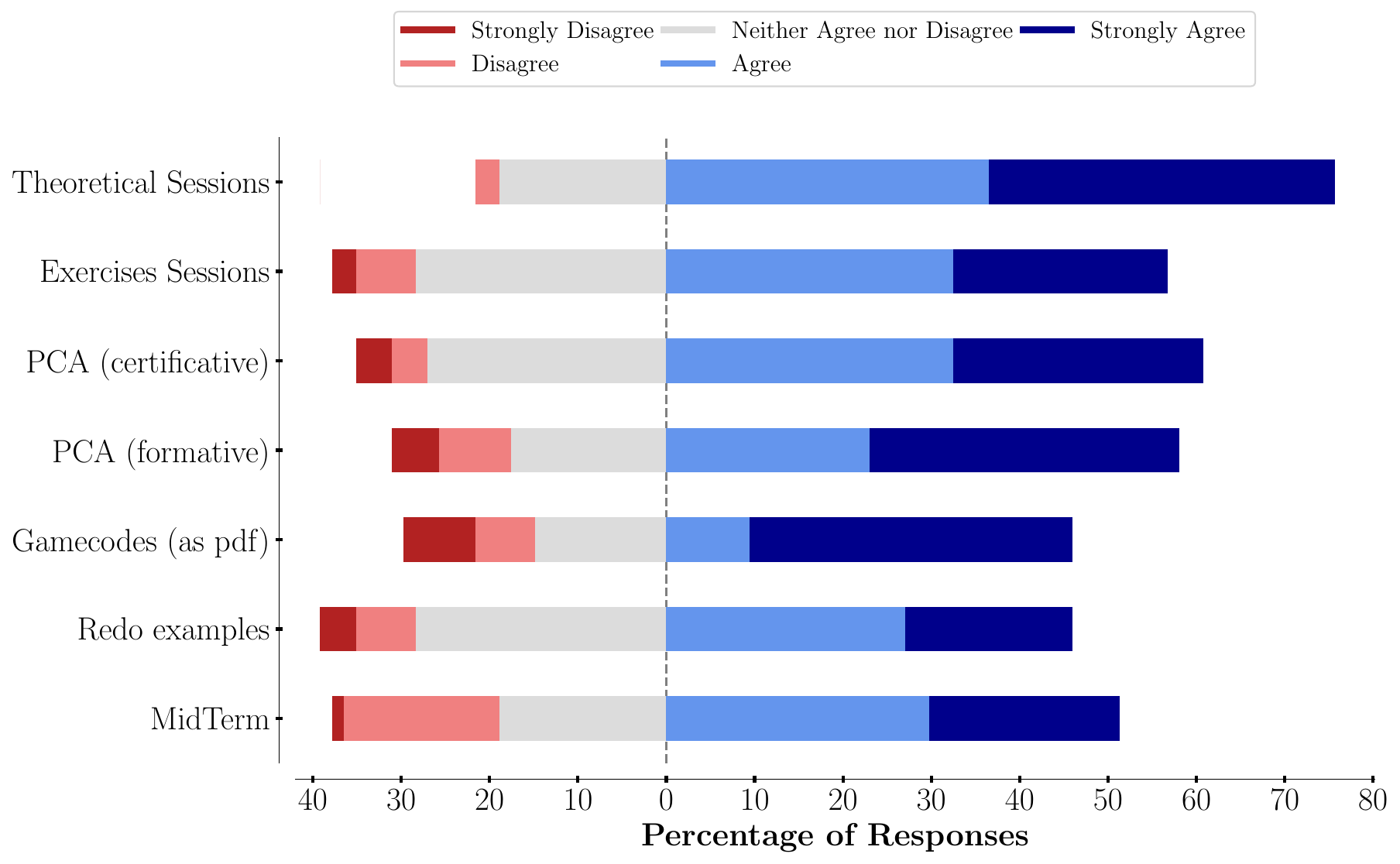}
  \end{center}
  \caption{Perception of the \pca with respect to other activities.}
  \label{cafe.percAct.fig}
\end{figure}

%%%%%%%%%%%%%%%%%%%%%%%%%%
This paper addresses a particular focus on three questions students were asked in the survey. The first question compares the impact of \cafe's experience on students' motivation in learning, with respect to other (more classic) activities. Fig.~\ref{cafe.percAct.fig} shows that the \pca (supported by \cafe and described in Sec.~\ref{pca}) comes as the second most stimulating activity, after the theoretical lessons. In particular, 60\% of the respondents (strongly) agreed that the \pca was motivating, 25\% had no opinion, and 15\% did not embrace that online experience. More precisely, both the certificative and the formative periods seem relevant to practice the course. Taking a closer look, the opinions are slightly stronger regarding the formative period. Likely, on the one hand, some students were more autonomous, allowing them to take benefit from the \pca independently from any grading pressure while others did not take the formative periods as an opportunity to train. This is also underpinned by Fig.~\ref{cafe.partic.access} (detailed below). One reason explaining some lower activity outside the certificative period could be that students got stuck despite the \fb, preventing them from progressing and achieving the \progchall. %To overcome that, the \gamecodes represent a good alternative since they offer detailed tips and theoretical reminders across students' resolution. Fig.~\ref{cafe.percAct.fig} illustrates that they had a very positive impact on 35\% of students. Although, another 35\% of students were more mitigated or didn't even share their opinion because they didn't experiment them. A possible reason to that is that the \gamecodes appeared too heavy as they were proposed through pdf files that semester.

\begin{figure}[!t]
  \begin{center}
      \includegraphics[width=8cm]{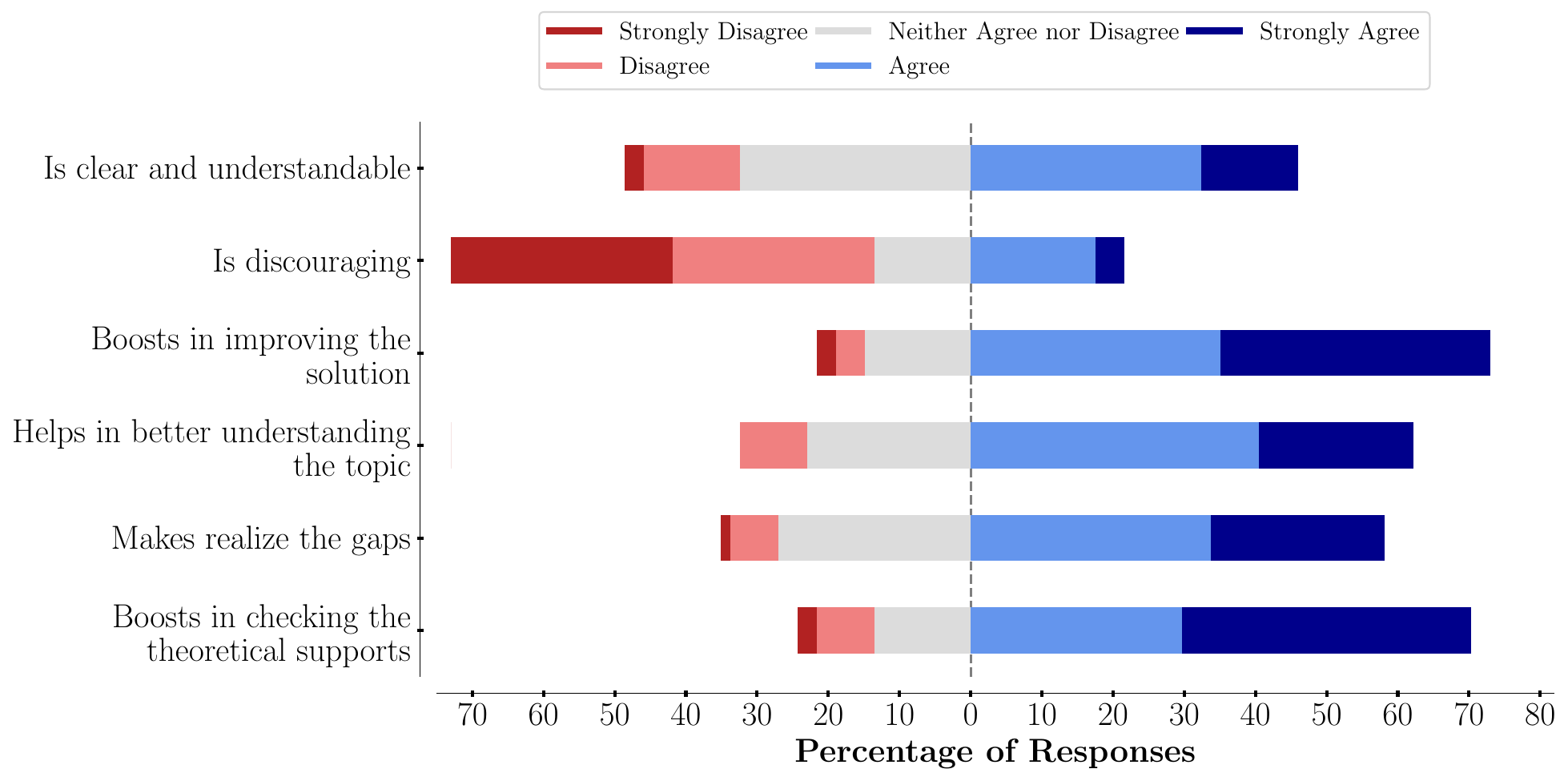}
  \end{center}
  \caption{Perception of the feedback.}
  \label{cafe.percFB.fig}
\end{figure}

Besides this, a special interest was dedicated to the way students perceive the automated \fb. From a general point of view, Fig.~\ref{cafe.percFB.fig} reflects that the \fb is well-received by the students. More precisely, half of the respondents found is clear and understandable, 30\% felt mitigated about it, and 20\% could not understand it well. Despite some misunderstanding of the \fb, a majority of respondents (74\%) felt boosted in improving their solution after receiving the \fb. In the same way, more than 60\% of respondents could identify their gaps, focus on the corresponding theoretical supports to fill them, and better understand the topic. Lastly, it is also interesting to note that, although some students were struggling in digesting the \fb, few of them really felt discouraged.

\begin{figure}[!t]
  \begin{center}
      \includegraphics[width=8cm]{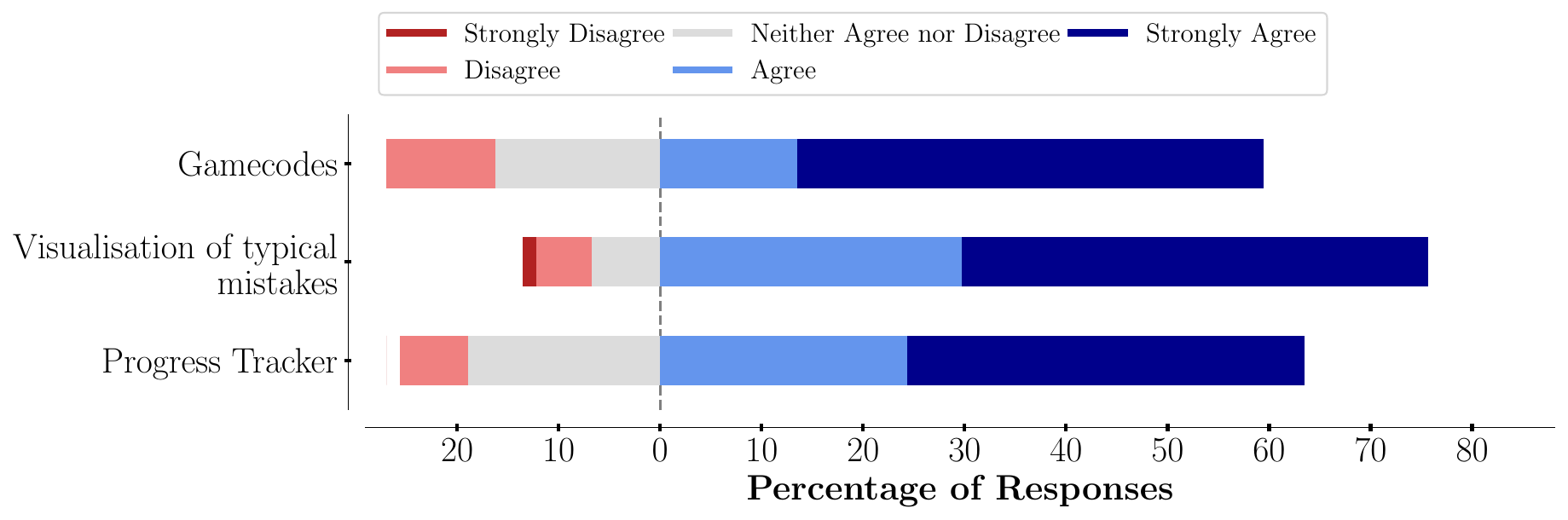}
  \end{center}
  \caption{Need for new functionalities.}
  \label{cafe.percNewFunc.fig}
\end{figure}

Finally, students were asked about how relevant it would be to integrate some new functionalities in \cafe. From Fig.~\ref{cafe.percNewFunc.fig}, it can be noticed that a large amount of respondents (78\%) would like to get more transparency regarding the mistakes they make. It suggests that students may miss self-assessment skills~\cite{difficultSelfAssessment}. In the same vein, many students would appreciate to visualize their actual progress with respect to what has been taught in classroom activities so far. It may help them in self-regulating their time by being aware of where their actually stand with respect to the course expectations. Next, 45\% of respondents expressed they strongly need the \gamecodes to be integrated in \cafe, which is fully consistent with the proportion of students whose learning got boosted by the \gamecodes (see Fig.~\ref{cafe.percAct.fig}). That need of digitalising the \gamecodes makes sense as, currently, the \gamecodes are presented as interactive PDF files where the different pages represent the resolution step, which results in heavy documents and low quality-of-experience. Due to its dynamics nature, the \gamecodes should definitely be transposed in an only plateform where the content can be organised so that the student easily access the information they need, the rest being fully hidden.
%\ed{GB : A voir si on garde ou pas. Si on met ceci en perspective avec les autres fonctionnalités proposées, on voit que les étudiants sont un peu demandeurs de tout en fait}

\subsubsection{Participation}\label{eval.results.participation}
\begin{figure*}[!t]
  \begin{center}
    \subfloat[Participation to the \progchalls.]{
      \label{cafe.partic.fig}
      \includegraphics[width=0.4\textwidth]{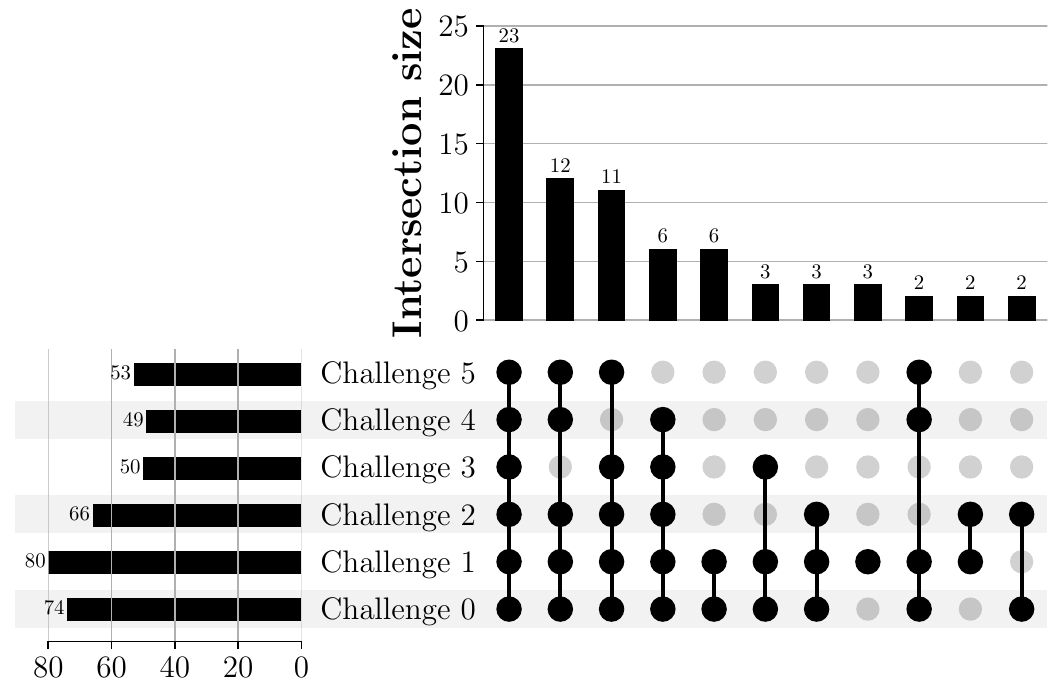}
    }
    \subfloat[Accessing \cafe during Academic Year 2022--2023.]{
      \label{cafe.partic.access}
      \includegraphics[width=0.6\textwidth]{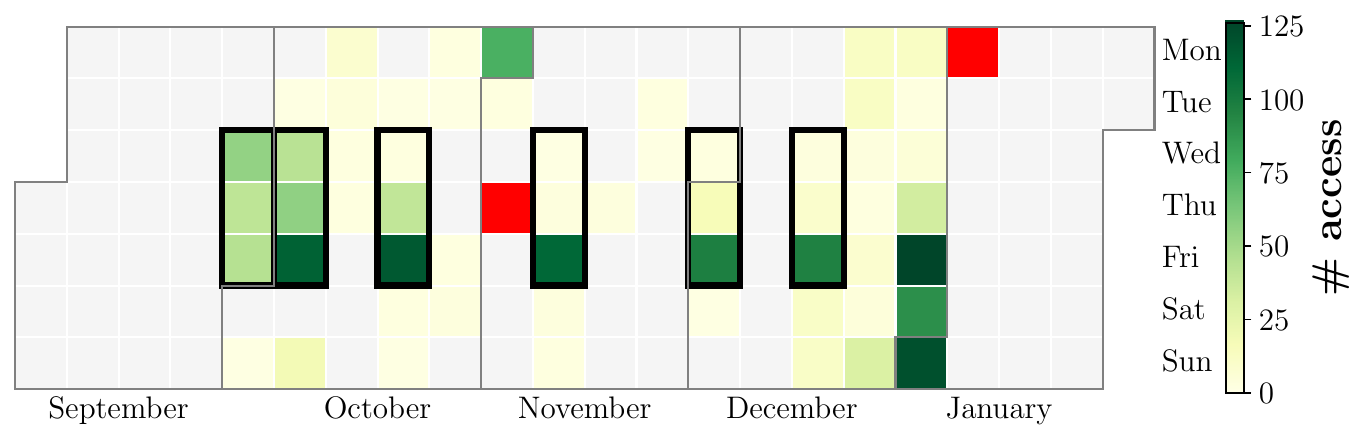}
    }
  \end{center}
  \caption{Results on participation.}
  \label{eval.results.participation.fig}
\end{figure*}

%%%%%%%%%%%%%%%%%%%%%%%%%%%%%%
%Now we have captured students' perception regarding \cafe, we can shift our attention to their actual participation on that learning plateform.
Fig.~\ref{cafe.partic.fig} depicts an upset plot~\cite{upset} illustrating how students participated to the \pca. The figure is made up of three parts: the matrix (bottom right) shows to participation. A dot in the matrix means that at least one student has participated to that \progchall. If there are multiple black dots on a column, it corresponds to students having participated to multiple \progchalls (e.g., the first column refers to students having participated to the six \progchalls). The histogram on the bottom left is the size of each matrix row, while the histogram on top right gives the number of students in the corresponding column of the matrix. We see that \progchall 1 was the most carried out (82\% of registered students) but the number of students dropped over time. This is normal in our country as we face a large attrition rate during the first semester for new comers. Only 23 students took the six \progchalls (23.8\% of registered students).

Then, Fig~\ref{cafe.partic.access} distills students' participation over time by highlighting, for each day of the semester, the number of students who started a session on \cafe. More specifically, the two red squares mark the two main evaluations of the course : the midterm and the final exam. Next, the black rectangles refer to the certificative periods of the \pca. Knowing that, we can note that the number of sessions consistently concentrate during those periods and just before the exam. We can also notice a higher number of sessions a few days before the midterm. The days in-between were dedicated to midterms related to other courses, that is likely why few or no session got launched those days. More generally, it appears that students get the most active only when they are against some evaluation rather than regularly practicing to master what they are being taught. That statement corroborates many research results \cite{DifficultyselfRegulate, DifficultyselfRegulate1, lessonLearned} stating that students have difficulties to self-regulate. However, we can still notice few connections spread along the semester, reflecting students who autonomously took benefits from \cafe.

%How long students stay on the feedback? Not easy to be sure about this data as session can remain open on the feedback even if there's no activity from the student

% !TEX root = ./paper.tex
\section{Extending \cafe}\label{extension}
%%%%%%%%%%%%%%%%%%%%%%%%%
In Sec.~\ref{cafe}, \cafe was introduced as an interdisciplinary learning tool, aiming to train abstraction and problem-solving skills in general. This section consolidates that ambition by detailing the preparation to perform in order to fit \cafe with a new problem profile, related to a STEM field. %whatever the field, as long as its purpose is to solve problems.

\subsection{General Consideration}\label{extension.general}
%%%%%%%%%%%%%%%%%%%%%%%%%%%%%%%%%
%Focussing on each module described in Sec.\ref{cafe}, here is what should be handled beforehand to instantiate \cafe for a new discipline : \cafe may match with any field whose purpose is to solve problems.  For a given field (Maths, Physics, Computer Science ...), different problem profiles can be identified.
To integrate a new problem profile in \cafe, the following requirements must be met :
\begin{enumerate}[label=Requirement \arabic*, align=left]
  \item : The resolution should be paved by a sequence of production steps.\label{req.sequence}
  \item : The resolution should run through two phases : the abstraction one and the concrete one.\label{req.abs}
  \item : The abstraction phase should rely on a graphical reasoning.\label{req.drawing}
  \item : The graphical representation should be dynamic, in such a way that it can be manipulated to illustrate different solution states (general ones and specific ones).\label{req.manip}
  \item : The graphical representation should be made up of predefined graphical components. They can stand as placeholders or movable elements students must handle when they are designing a solution. \label{req.graphComp}

\end{enumerate}

Besides this, independently from the discipline, an activity (that can be similar to the \pca, described in Sec.~\ref{pca}) requires to be set up to make \cafe standing as an integral part of the whole course activities.  Finally, it is worth noticing that having first year students as target public is the most appropriate in order to avoid too complex solution modeling. Moreover, it is likely that first year students are those who need this kind of support the most.

\subsection{Application to Physics}\label{extension.physics}
%%%%%%%%%%%%%%%%%%%%%%%%%%%%%%%%%%%
To emphase the interdisciplinary potential of \cafe, we match it to a specific problem profile picked from another field than Computer Science. In particular, we are considering the following Kinematics problem in Physics : \\

\fbox{\begin{minipage}{23em}
A car of mass $m=1200kg$ is parked on a slope of $\alpha=30^\circ$. We would like to compute the magnitude of the friction forces so that the car is at rest.
\end{minipage}} \\

\begin{figure*}[!t]
  \begin{center}
    \subfloat[Resolution Flow in Physics]{
      \label{cafe.physResol.fig}
      \includegraphics[width=1.2\columnwidth]{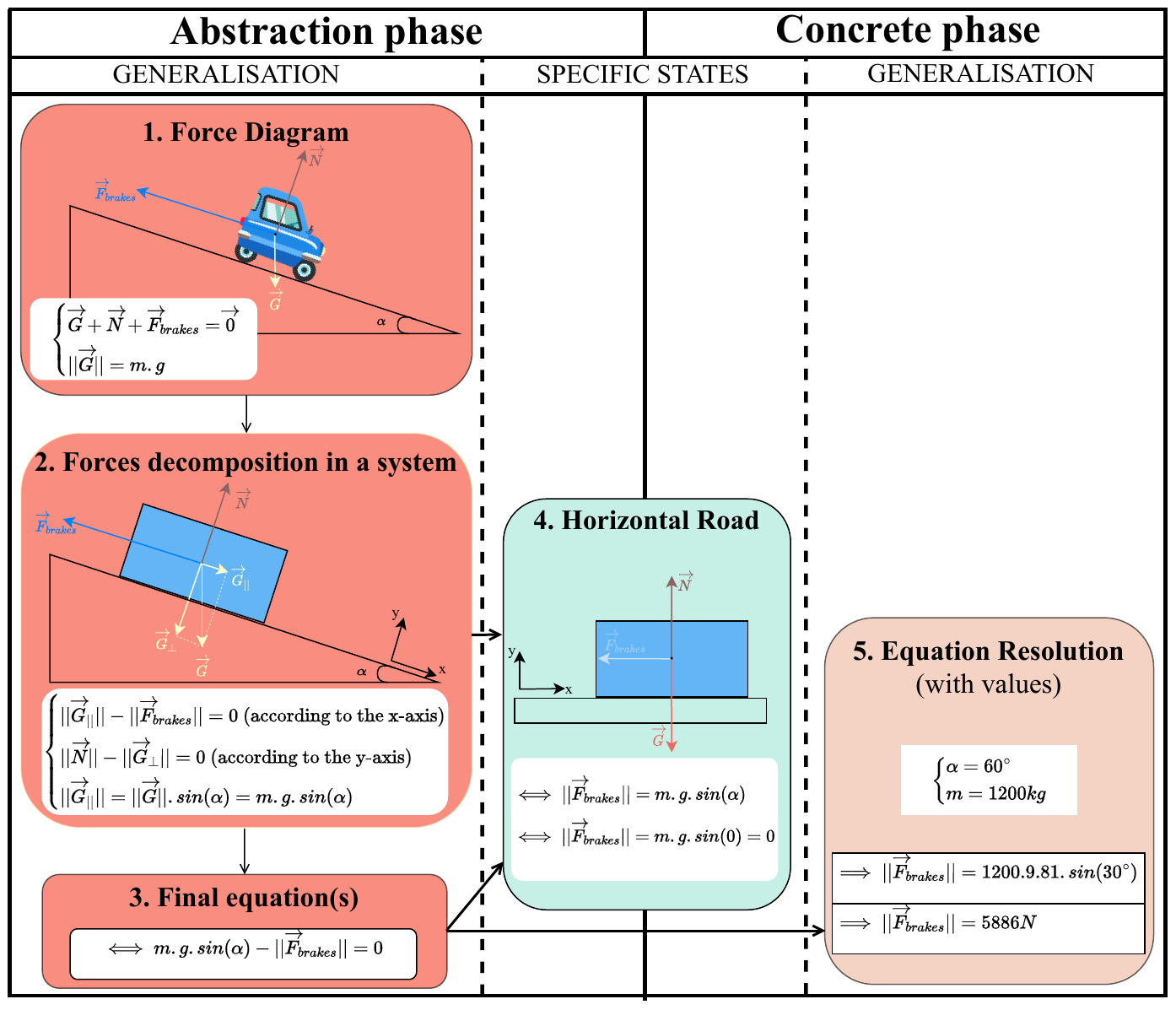}
    }
    \subfloat[Blank Diagram in Physics.]{
      \label{cafe.physBlank.fig}
      \includegraphics[width=0.8\columnwidth]{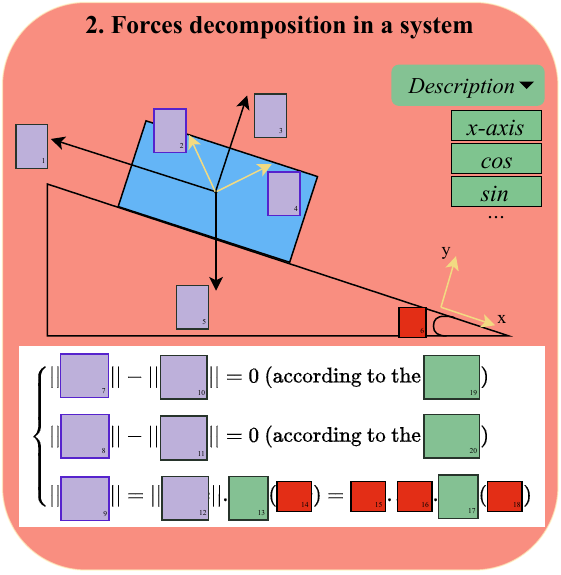}
    }
  \end{center}
  \caption{How \cafe can fit to a Physics Introduction course.}
  \label{cafe.phys.fig}
\end{figure*}

\subsubsection{Activity Resolution Setting}\label{extension.activity_resolution}
%As presented in the previous subsection and through Fig.~\ref{cafe.actGen.resol.fig}, the problem resolution should rely on some predefined sequential productions. Furthermore, each production should be incorporated in one of the abstraction or the concrete phases (or bridging them).
Considering that specific problem type, five productions could be defined : ($i$) Representing the situation and identifying the forces that are applied on the object of interest (being the car here) ; ($ii$) Choosing a system and decomposing the different forces so that they follow the direction imposed by the system ; ($iii$) Deriving the mathematical expression(s) allowing one to formulate the friction forces ; ($iv$) Transposing the general representation in a particular case (in which the problem might be possible to be intuitively solved) ; ($v$) Using numerical data to compute the solution. Fig.~\ref{cafe.physResol.fig} illustrates those productions and map them to the two phases \cafe is supporting (\ref{req.abs}). Also, Fig.~\ref{cafe.physResol.fig} shows that the resolution is done in sequence (\ref{req.sequence}).  Further, the Force Diagram corresponds to \ref{req.drawing}. Finally, that drawing is manipulated on step 4, corresponding to \ref{req.manip}.

\subsubsection{Correction and \Fb Setting}\label{extension.correction}
%%%%%%%%%%%%%%%%%%%%%%%%%%%%%%%%%%%%%%%%
To enable automatic Correction and \Fb, a \misconception needs to be defined and fed. That library should cover most of the mistakes students may fall in across their resolution, as explained in Sec~\ref{cafe.correction} and illustrated in Fig.~\ref{cafe.FB.fig}. For instance, considering our Kinematics problem, for the second production where forces should be decomposed according to the system, one typical mistake might be that the students did not direct all the arrows according to the orthonormed system that is set. Another example of mistake could be that the second equation does not reflect the drawing above.

\subsubsection{Production Modeling Setting}\label{extension.blank_drawing}
%%%%%%%%%%%%%%%%%%%%%%%%%%%%%%%%%%%%%%%%%
Similarly to the \bgli, it is relevant to hide from the expected pictorial representation of the situation the key components. In this way, students must identify and link them on their own while still having benchmarks thanks to the provided canvas. Furthermore, on the supervisor's side, those components need to be modeled (by having a specific semantic and relation with each other) in order to enable automatic personalized feedback. For the Kinematics problem of interest, a relevant blank diagram could be the one exposed through Fig.~\ref{cafe.physBlank.fig}. That figure relies on different kind of graphical components, each of them being mapped to a specific color. Like in the \bgli, red boxes should host variables while green ones expect some description picked from a dropdown list. In addition to them, purple boxes stand for forces and movable arrows are illustrated in yellow. All the rest is fixed.

\subsubsection{Drawing Editor Setting}\label{extension.drawing_editor}
%%%%%%%%%%%%%%%%%%%%%%%%%%%%%%%%%%%%
As suggested above, the graphical components that would be useful to formally represent a situation in Kinematics (\ref{req.graphComp}) are variables, forces, predefined descriptions, and movable arrows. In addition to them, some relevant patterns should be identified and designed. Then, they should be integrated in the Drawing Editor so that, once a teacher or a student wants to depict a scenario in Kinematics, they can simply select a suitable pattern and drag and drop the graphical components to build up their pictorial representation.

% !TEX root = ./paper.tex
\section{Related Work}\label{related}
%%%%%%%%%%%%%%%%%%%%%%

%Automated System with Automatic Correction
Many automated system providing programming exercises were already proposed (e.g.,~\cite{webcat, problets, inginious, codingbat, myproglab, unlock,coderunner}).  Most of them apply test-based correction, i.e., student's code is corrected through unit testing (except \dfn{UNLOCK}~\cite{unlock} that tackles the problem solving skills in general, not just coding skills). \dfn{WebCAT}~\cite{webcat} even makes students write their own tests too. \textit{Kumar's Problets}~\cite{problets} enables step by step code execution as part of the \fb. Closer to \cafe, \dfn{Dodona}~\cite{tool_dodona} proposes programming assignments with automated \fb and harness the data to regulate the teaching materials. However, Dodona is specialized in practicing coding (considering different programming languages) in a collaborative environment (by allowing students to ask questions on a forum) while \cafe focuses on students' abstract thinking, upstream to their code. Other tools also offer some features depicting an abstract representation of some pieces of code. Among them, \dfn{Virtual-C IDE}~\cite{tool_vide} typically allows to visualize a C-program behavior. \dfn{Online Python tutor}~\cite{python_tutor} allows students to execute their code step-by-step and visualize the runtime state of each structure. And \dfn{JASM.IN}~\cite{jasmin} illustrates dynamic program state and provides state transition rules for the execution of Java language constructs. \cafe differentiates from those three tools by guiding students in modeling their solution by themselves, so that they can rely on it to build their code. On the contrary, Virtual-C IDE and JASM.IN automatically post-represent the solution based on some pieces of code students have submitted. Considering now the tools where students are designing their solution on their own, you can namely find \dfn{Python turtle graphic library} that was already demonstrated as improving abstraction skills~\cite{python_abstraction}. However, \cafe goes further than that as it bridges students' graphical representation and resulting code by checking the consistency between those two versions of the solution and provides automatic \fb with respect to that.

%Automatic Feedback
Automatic \fb has been extensively motivated, described, and discussed in previous studies~\cite{feedback_survey1, feedback_survey2}.  In particular, Keuning et al.~\cite{feedback_survey2} have shown that it is not that easy to tune \fb with respect to some specific needs (proper to the topic and the statement). If we position \cafe with respect to \fb theoretical aspects, \cafe implements ``Answer-until-correct'' (AUC)~\cite{narciss} \fb, as students can refresh their solution as much as they need to.  It is similar to Singh et al.~\cite{feedback_singh} where students get a numerical value (the number of required changes) and the suggestion(s) on how to correct the mistake(s).  In addition, \cafe meets the principles Nicol introduced around student's engagement, self-regulation, and academic experience (the only missing dimension being the social one)~\cite{nicol}.

% Discipline
Finally, regarding the methodology \cafe is currently supporting, other studies promoted problem-solving through predefined steps~\cite{solutionThroughSteps}. Furthermore, the relevance of transiting through an abstract representation of a solution was corroborated to prevent students from getting overwhelmed with specific problem instances~\cite{importanceAbstractRepres}.

% Loop Invariant and Graphical Reasoning - With respect to graphical reasoning,  Tam~\cite{teaching_loop_invariant} suggests to introduce students to \invariant as early as possible in their cursus and describes several examples of code construction based on informal \invariants expressed in natural language. Astrachan~\cite{invariants_pictures} suggests the use of \glis in the context of CS1/CS2 courses.  However, his approach is incomplete as the suggested drawing lack of completeness (e.g., objects manipulated, such as arrays, are not named in the drawing), might lead to confusion (e.g., variables positions around the dividing line are somewhat unclear), and the drawing is not explicitly manipulated to derive particular situations.  Back~\cite{ibp_back,back,back2} proposes nested diagrams (a kind of state charts) representing, at the same time, the \invariant and the code. However, in such a situation, \invariants are expressed as logical assertions. Since, Manilla~\cite{invariant_edu} has evaluated the impact of errors in those nested diagrams.   Finally, Erkisson et al.~\cite{pictural_inv} propose a pictural language for representing \invariants.  Their language only applies to arrays and is a mix between drawings (the data structure is drawn and partitions are colored to illustrate universally quantified predicate) and formal languages (the meaning of partitions is expressed as a predicate). None of them have provided students with a tool helping to draw a \gli and to manipulate it to derive the code.

% !TEX root = ./paper.tex
\section{Conclusion}\label{ccl}
%%%%%%%%%%%%%%%%%%%%%%%%%%%%%%%
This paper describes \cafe, a tool whose purpose is to make students regularly and activily work in order to maintain them on a correct track, by providing instantaneous personalized \fb. In practice, \cafe supports online activities that are made up of problems to solve. For each new statement, besides its definition, its corresponding solution needs to be outlined and configured by a supervisor. In particular, the solution should be articulated by different types of productions, each of them being framed through a canvas with specific placeholders whose semantic and potential resulting typical mistakes should be parametrized beforehand. This way, on the one hand, students get guided across their resolution and, on the other hand, automatic and personalized review gets enabled since student's solutions can be anticipated.

\cafe is relevant to be integrated in any course where abstract representation through drawings stands as the ground for constructing a solution. More specifically, in our CS1 course, abstraction is introduced in the context of building a loop piece of code. More precisely, the \glibpFull methodology is taught. Therefore, the central kind of canvas in a resolution flow is the \bgli, representing the shape of a \gli with different kinds of boxes to fill and movable bars, allowing to visualize the solution under construction at a specific iteration.

As further work, \cafe could be transposed in other disciplines dedicated to first year students. The first steps towards such an extension have been already taken in the context of a Physics course.

Besides this, more specifically, \cafe's set of functionalities keeps being expanded in terms of activities and data processing. First, regarding the activities, in addition to the \pca, the \gamecodes (Sec.~\ref{activity}) are under implementation. Next, on the long run, it is aimed to implement a third kind of activity : the \cdb activity~\cite{cdb}. The purpose of this activity drastically differs from the \pca and the \gamecode as it targets to motivate our programming methodology by applying it in a real-life scenario rather than academically training it. Despite that differentiation, it remains very relevant to integrate such an activity in \cafe as it also relies on a sequence of productions to submit, in response to a given problem. The only difference is that, for each production, students can get more freedom since the resulting ``instantaneous'' \fb is built by students reviewers rather than a machine~\cite{cdb2}. Therefore, the difficulty in putting in place such an activity is shifted from the automatic \fb configuration to the peer-\fb reviewing process where teams should be defined, time should be punctuated, \fb should be supervised, and productions should move between students. Besides the activities, it is also planned to dedicate a focus on the Progress Tracker and the Learning Analytics Dashboard in order to get more transparency about students' learning activity and wisely adapt the content students should practice.

%\newpage
\bibliographystyle{IEEEtran}
\bibliography{Bibliography}

% !TEX root = ./paper.tex
\vspace{-1cm}
\begin{IEEEbiography}
  [{\includegraphics[width=0.9in,height=1.125in,clip,keepaspectratio]
  {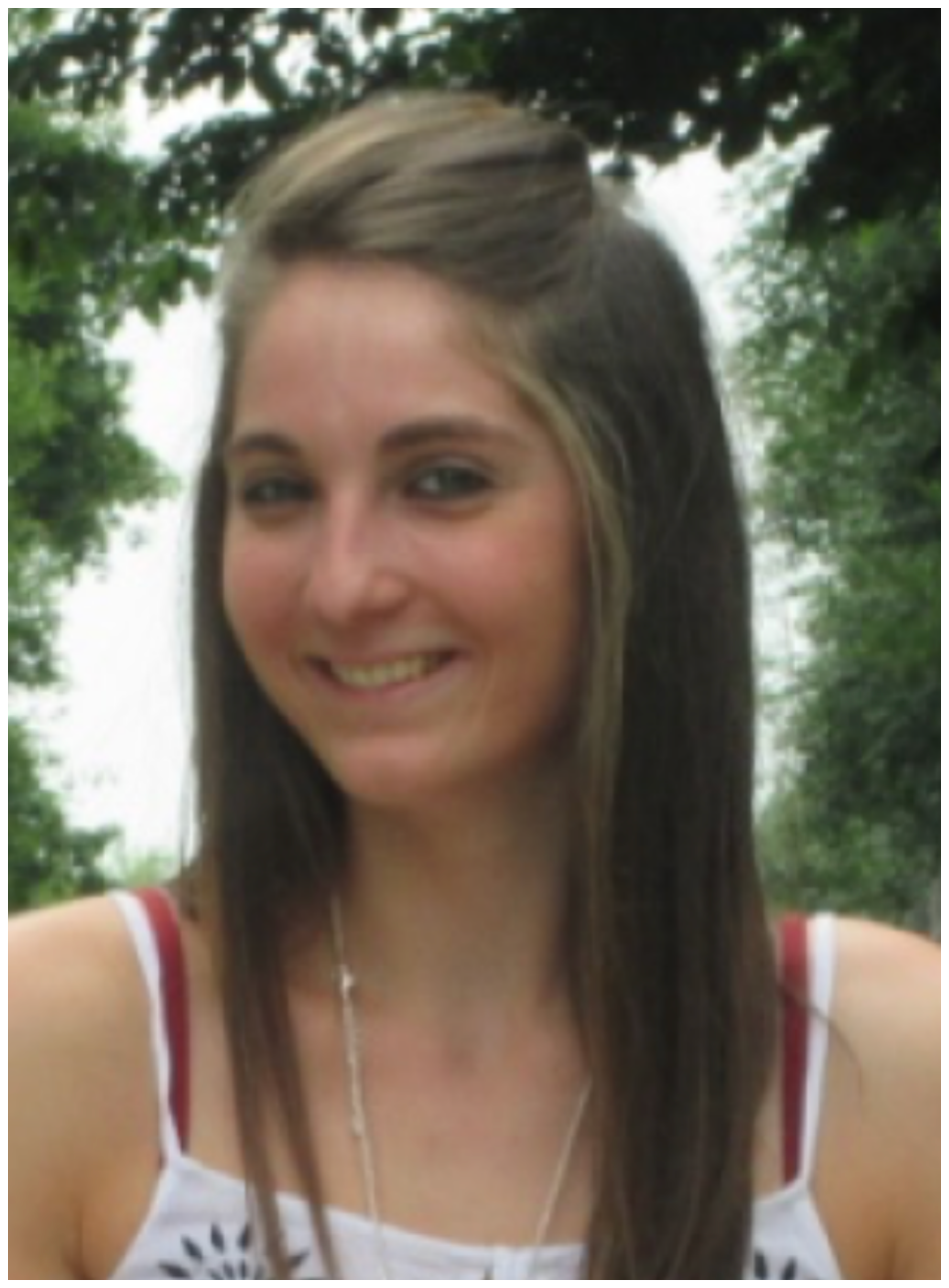}}]{G\'eraldine Brieven} is a Ph.D. student in Computer Science Education at the University of Li\`ege since 2021. She got a master's degree in Civil Engineering in Computer Science at the same University, in 2019. Before starting her Ph.D., she worked for two years in an IT Consultancy company. Her research interests are about Collaborative Learning and Computer-Assisted Learning (relying on Automatic Feedback). It aims to motivate and train problem-solving skills. % as well as a Teaching Assistant 
\end{IEEEbiography}
\vspace{-1.5cm}
\begin{IEEEbiography}
  [{\includegraphics[width=0.9in,height=1.125in,clip,keepaspectratio]
  {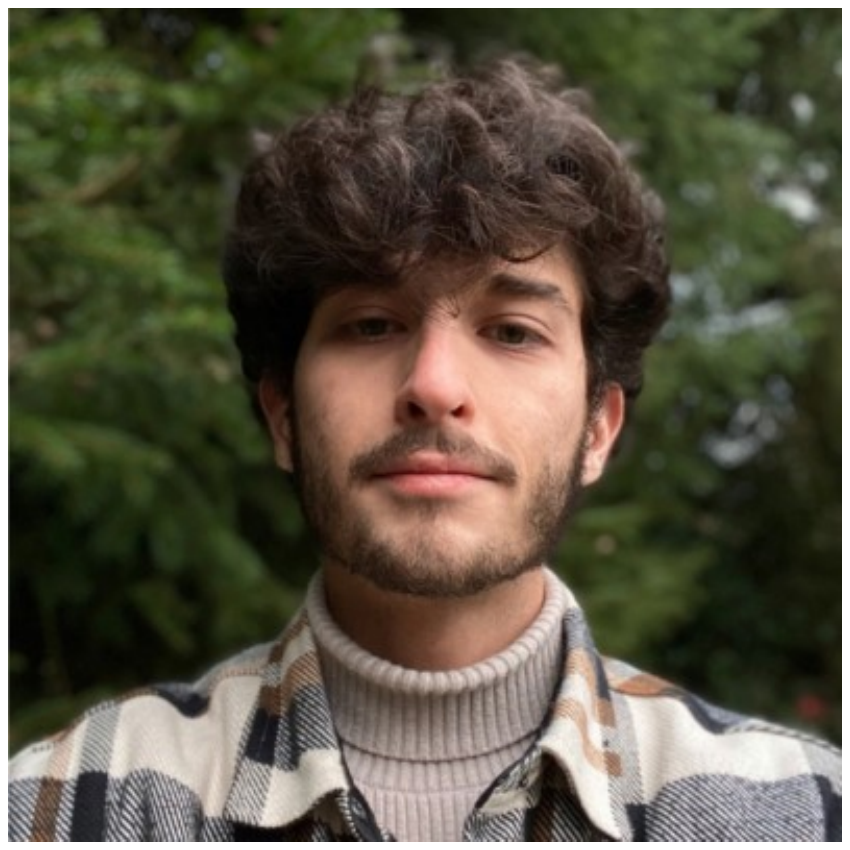}}]{Lev Malcev}  finished his bachelor's degree in Computer Science at the University of Li\`ege in 2021. He is currently working as a freelance developer and finishing his master's degree in Computer Security. He is also involved in developing \cafe, a tool giving students the opportunity to train their problem-solving skills and allowing to collect learning analytics to study students' learning behavior.
\end{IEEEbiography}
\vspace{-1.5cm}
\begin{IEEEbiography}
  [{\includegraphics[width=0.9in,height=1.125in,clip,keepaspectratio]
  {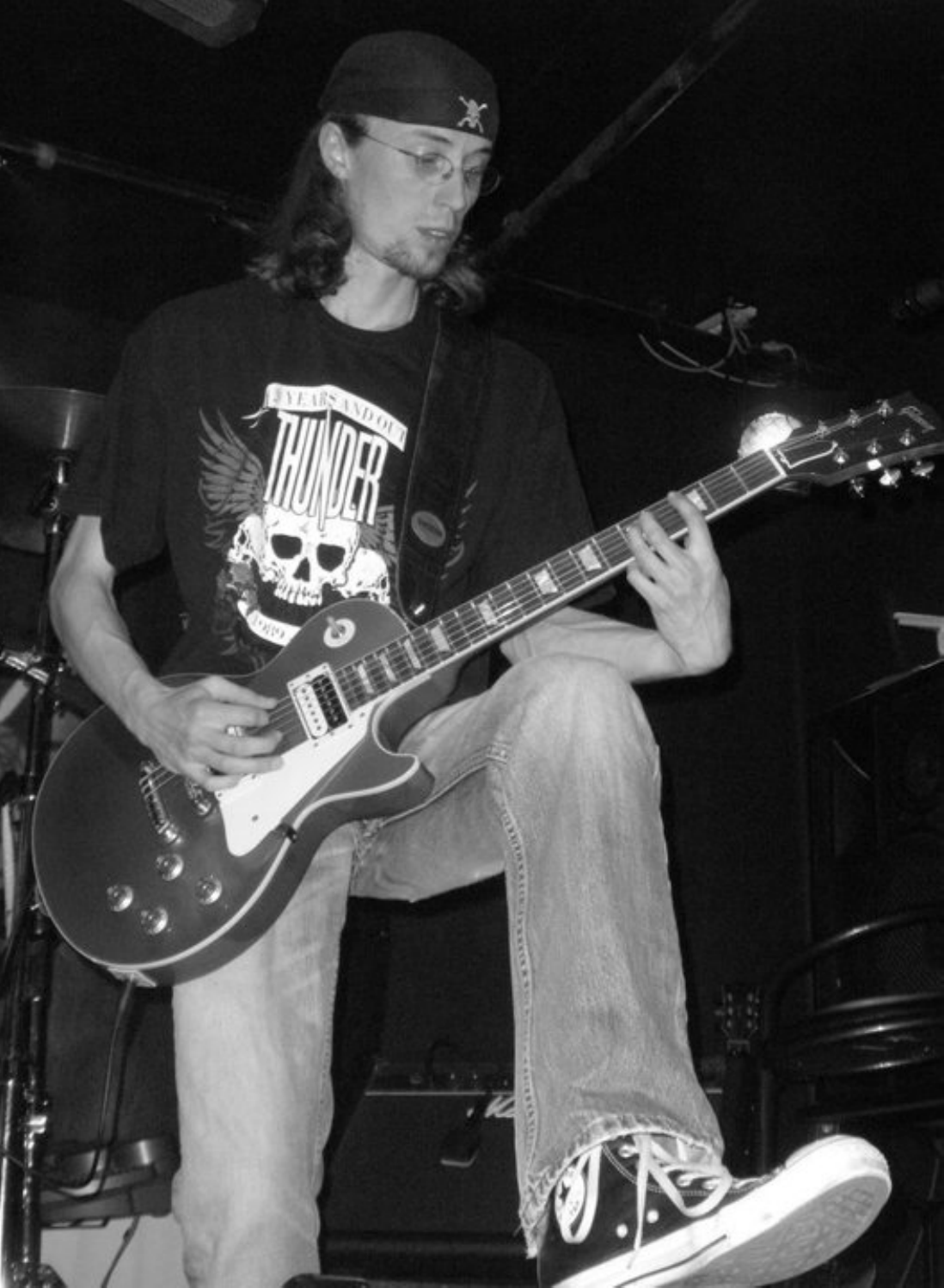}}]{Benoit Donnet} received his Ph.D. degree in Computer Science from the Universit\'e Pierre et Marie Curie in 2006 and has been a PostDoc until 2011 at the Universit\'e catholique de Louvain (Belgium). Mr.~Donnet joined the Montefiore Institute at the Universit\'e de Li\`ege since 2011 where he was appointed successively as Assistant Professor and Associate Professor. His research interests are about Internet measurements (measurements scalability, Internet topology discovery, measurements applied to security), network modeling, middleboxes, new Internet architectures (LISP, Segment Routing), and Computer Science Education.
\end{IEEEbiography}

\end{document}